\documentclass[a4paper,11pt]{article}
\pdfoutput=1 
\usepackage{jheppub} 
\usepackage[T1]{fontenc} 
\usepackage{epsfig}
\usepackage{caption}
\usepackage{subcaption}
\usepackage{color,soul}
\usepackage{float}
\usepackage{url}
\usepackage{slashed}
\usepackage{accents}
\usepackage[normalem]{ulem}
\def\beq{\begin{equation}}
\def\eeq{\end{equation}}
\def\beqn{\begin{eqnarray}}
\def\eeqn{\end{eqnarray}}

\def\gev{{\rm~GeV}}
\def\tev{{\rm~TeV}}

\title{Higgs Couplings at High Scales}

\author[a]{Dorival Gon\c{c}alves,}
\author[a]{Tao Han}
\author[a,b]{and Satyanarayan Mukhopadhyay}

\affiliation[a]{PITT-PACC, Department of Physics and Astronomy, University of Pittsburgh, PA 15260, USA}
\affiliation[b]{Department of Theoretical Physics, Indian Association for the Cultivation of Science, Kolkata 700032, India}

\emailAdd{dorival@pitt.edu}
\emailAdd{than@pitt.edu}
\emailAdd{tpsnm@iacs.res.in}

\preprint{PITT-PACC-1801}
\abstract{
We study the off-shell production of the Higgs boson at the LHC to probe Higgs physics at higher energy scales utilizing the process $g g \rightarrow h^{*} \rightarrow ZZ$. We focus on the energy scale-dependence of the off-shell Higgs propagation, and of the top quark Yukawa coupling, $y_t (Q^2)$. Extending our recent study in~\cite{Goncalves:2017iub}, we first discuss threshold effects  in the Higgs propagator due to the existence of new states, such as a gauge singlet scalar portal, and a possible continuum of states in a conformal limit, both of which would be difficult to discover in other traditional searches. We then examine the modification of $y_t (Q^2)$ from its Standard Model (SM) prediction in terms of the renormalization group running of the top Yukawa, which could be significant in the presence of large flat extra-dimensions. Finally, we explore possible strongly coupled new physics in the top-Higgs sector that can lead to the appearance of a non-local $Q^2$-dependent form factor in the effective top-Higgs vertex. We find that considerable deviations compared to the SM prediction in the invariant mass distribution of the $Z$-boson pair can be conceivable, and may be probed at a $2\sigma$-level at the high-luminosity 14 TeV HL-LHC for a new physics scale up to $\mathcal{O}(1 {~\rm TeV})$, and at the upgraded 27 TeV HE-LHC for a scale up to $\mathcal{O}(3 {~\rm TeV})$. For a few favorable scenarios, $5\sigma$-level observation may be possible at the HE-LHC for a scale of about $\mathcal{O}(1 {~\rm TeV})$.
}

\begin{document} 
\maketitle
\flushbottom

\section{Introduction}

With the milestone discovery of the Higgs boson at the CERN Large Hadron Collider (LHC) \cite{Observations}, elementary particle physics has entered a new era: For the first time ever, we have a self-consistent, relativistic and quantum-mechanical theory, the Standard Model (SM), that could be valid all the way to an exponentially high scale. However, the SM Higgs boson possesses a profound puzzle: While the Higgs mechanism provides the mass of all elementary particles in the SM by their couplings ($g_i$) to the Higgs doublet, $m_i \propto g_i v$, where $v$ is the vacuum expectation value of the Higgs field, it is not clear how the Higgs boson mass itself can be stabilized at the electroweak scale against quantum corrections from new physics beyond the SM. The quadratic sensitivity of its mass correction to the new physics scale $\Lambda$, $\delta m_h^2 \propto g_i^2 \Lambda^2$, is commonly viewed as a hint for the existence of physics beyond the SM (BSM) not far from the electroweak scale \cite{Naturalness}, such as the constructions of weak-scale supersymmetry \cite{susy} or strong dynamics of composite Higgs \cite{strong,Csaki:2015hcd}.

Well-motivated experimental efforts have been carried out in the search for BSM physics at the TeV scale associated with the naturalness argument. However, all the searches so far have led to null results. In the absence of new physics signals, especially in the experiments from the LHC at the energy frontier, it is conceivable that the solutions to the naturalness puzzle might have taken a more subtle incarnation, not captured by the usual signatures from the partners of the top quark, Higgs boson and gauge bosons in supersymmetry or with new strong dynamics in the Higgs sector. Thus the new states are either more difficult to observe, or our notion of naturalness based on the Wilsonian paradigm of effective field theory should be revisited. In both these scenarios, new search strategies would have to be developed to uncover the underlying dynamics or principles associated with the electroweak sector. 

With these alternatives in mind, we would like to argue that it may be most rewarding to study the Higgs boson couplings at higher energy scales. If the new states responsible for naturalness are within the reach at the TeV scale, but hidden in the standard searches, they would necessarily show up in the energy scale-dependence of Higgs couplings, or more broadly in Higgs production processes through quantum corrections. If instead the additional states associated with the new dynamics in the Higgs sector are not within accessible energies, we could still expect deviations in Higgs processes at higher energies if the Higgs particle is non-elementary in nature, being a bound state of a new strong dynamics. 

To illustrate the first possibility, where the new physics scale is accessible but the new states are not readily observable by standard searches, we consider three  example scenarios, in the first two of which there are new effects in the energy scale dependence of the Higgs propagation, while in the third scenario the scale dependence appears in the top quark Yukawa coupling. As a first example, following our recent study~\cite{Goncalves:2017iub}, a new scalar singlet sector is introduced in the low-energy effective theory that couples to the Higgs field. Such a scalar sector can have implications for the little-hierarchy problem. We show that an effective way to probe these new states is to study their impact in the next-to-leading order electroweak corrections to Higgs production at higher energy scales. This is a minimal setup that illustrates the core idea behind our study. 

In the second example, we  discuss a rather striking scenario in which the Higgs sector approaches a conformal symmetry at high energies, suppressing the sensitivity to new physics in the Higgs mass corrections. An interesting example of this class is the quantum critical Higgs (QCH) \cite{Bellazzini:2015cgj}, where the Higgs properties are modified by a continuum spectrum coming from the conformal symmetry near the quantum phase transition. To effectively probe the quantum critical point, we need to uncover the momentum-dependence of the Higgs interactions near the phase transition.

In our third example, we ask a natural question that arises in the discussion of Higgs couplings as a function of the energy scale: how does the renormalization group (RG) evolution of Higgs couplings get modified in the presence of new states? A renormalizable four-dimensional quantum field theory predicts a simple logarithmic scaling of couplings with the energy scale $Q^2$. Such quantum corrections are expected to induce effects of the order $(g^2/16\pi^2)\log(Q^2/m_h^2)$, which amounts to a few percent modification of the couplings between $Q^2 \sim m_h^2$ and a TeV$^2$. However, in the presence of non-trivial new dynamics, the running of the couplings could be altered  dramatically. We illustrate this using the running of top quark Yukawa coupling $y_t$ in an extra-dimensional setting, with the SM gauge and/or matter fields propagating in the bulk~\cite{ED_RGE,Appelquist:2000nn,Appelquist:2001mj,5D,6D}. Summing over the multiple equally spaced Kaluza-Klein resonance thresholds 
(with the same couplings as in the SM), leads to an asymptotically power-law running of $y_t$~\cite{ED_RGE}, with interesting implications for Higgs production at higher energies. The particular flat extra-dimensional scenario adopted has no direct implication for naturalness, however, the framework is illustrative of how different the largest  Higgs coupling in the SM can become in the ultraviolet (UV) regime.

We finally consider the scenario in which the Higgs boson is a composite state at energy scales around a TeV. In our analysis, we do not necessarily refer to any low-energy semi-perturbative effective theory (such as the modern version of composite Higgs), nor to a technicolor-like description of the Higgs in terms of its constituents. Instead, we parametrize the effect of a finite-sized composite Higgs boson coupling with a generic form factor, and study its implication in Higgs processes where the Higgs particle itself is still the relevant degree of freedom. 

While all Higgs couplings should be examined as a function of the energy scale, arguably, the first targets are the couplings to heavier SM particles, namely, the top quark and the $W$ and $Z$ bosons. To this end, a particularly interesting proposal is to study the off-shell Higgs contribution to the  $gg \rightarrow ZZ$ process. The large interference between the Higgs induced amplitude and the gluon-fusion background component results in an appreciable  off-shell Higgs rate, thus making it feasible  
to study the Higgs couplings to top quarks and $Z$ bosons at different energy scales. As we shall see in the subsequent sections, this feature can be utilized to probe several BSM scenarios related to Higgs physics. This treatment also captures the feature of non-local, momentum-dependent top-Higgs interactions.

The rest of the paper is organized as follows. In Sec.~\ref{sec:process}, we discuss the off-shell Higgs production and decay processes at the LHC that constitute an optimal target to study Higgs couplings at high energy scales. Following the above discussion, in Sec. \ref{sec:portal}, \ref{sec:conformal} , \ref{sec:RGE} 
and \ref{sec:composite}, we describe the scalar singlet portal, Quantum Critical Higgs (QCH), RG evolution of Higgs couplings in an extra-dimensional setting and the form-factor description for a generic composite Higgs boson, respectively. In each case, we also discuss the implications of the searches at the high-luminosity phase of the 14 TeV LHC as well as the proposed 27 TeV HE-LHC upgrade. We conclude with a summary of our results in Sec.~\ref{sec:summary}.

\section{Higgs Couplings at High Energies: The $p p \rightarrow ZZ$ Process}
\label{sec:process}

The ATLAS and CMS collaborations have so far established a consistent picture of the Higgs boson couplings at the EW scale $Q^2\approx m^2_h$: 
to top quarks directly \cite{ATLAStth,CMStth} and indirectly \cite{ATLASsigma,CMSzzh}, 
and to $W^+W^-$ \cite{ATLASwwh,CMSwwh}, $ZZ$ \cite{ATLASsigma,CMSzzh}, 
$\tau\tau$ \cite{ATLAStau,CMStau}, and $b\bar{b}$ directly \cite{ATLASbb,CMSbb}.
Experiments at the LHC will continue to probe the Higgs sector both at the Higgs mass scale as well as at higher scales.
The obvious first target to study the scale-dependence is the top-quark Yukawa coupling: not only is it the largest Higgs coupling in the SM, thereby playing a major role in the hierarchy problem, it is also ubiquitous from the measurement point of view appearing in the leading Higgs production process. The next consideration would be the couplings with $W$ and $Z$ bosons at higher scales. However, we expect these to have a lesser sensitivity to new dynamics since, to a first approximation, they are governed by the well-tested gauge couplings.
\begin{figure}[t!]
\centering
\vspace{0.6cm}
  \includegraphics[width=.33\textwidth]{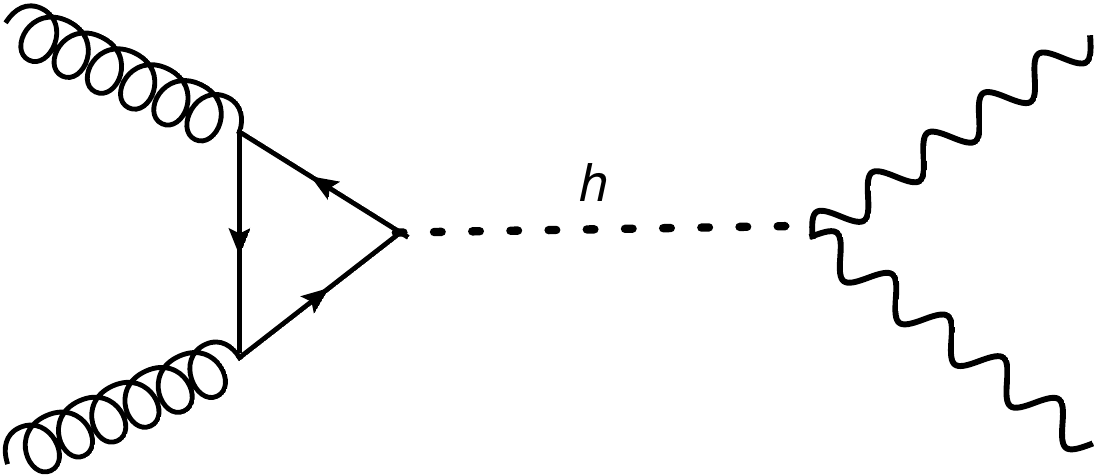}\hspace{2cm}
  \includegraphics[width=.27\textwidth]{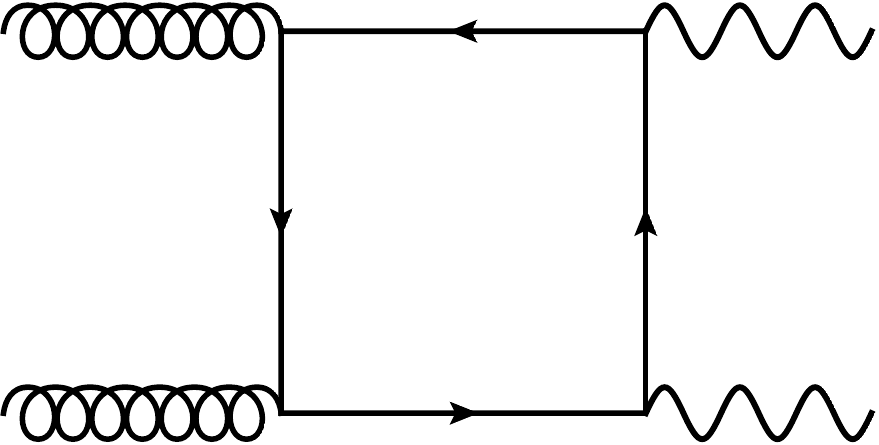}
 \caption{Representative  set of Feynman diagrams for $gg\rightarrow ZZ$ production in the SM :  involving the Higgs boson (left) and the SM fermion box diagram (right).}
 \label{fig:feyndiag1}
\end{figure}

In order to probe the energy-dependence of Higgs couplings, and more generally Higgs processes, we shall study the off-shell Higgs production via gluon fusion $g g \rightarrow ZZ$ at the LHC, a representative set of Feynman diagrams for which in the SM are shown in Fig.~\ref{fig:feyndiag1}. An important aspect of this process is that it presents three kinematical thresholds near $m_h, 2M_Z$ and $2m_t$ \cite{offshell}. These thresholds arise from the real part of the amplitudes from an $s$-channel resonance and the imaginary part near the pair-production thresholds.
It is encouraging to note that the event rate for this process at the LHC is substantial $-$ about one in every ten events for the process $ g g \rightarrow h^* \rightarrow Z Z$ involves an intermediate Higgs boson with invariant mass  $Q^2 > 4M_Z^2$~\cite{offshell}. Thus we will focus on the clean final state with four charged leptons
\begin{equation}
{pp\rightarrow h^* \to Z^{(*)}Z^{(*)} \rightarrow 4\ell}. 
\end{equation}

\begin{figure}[t!]
\centering
\vspace{0.6cm}
   \includegraphics[width=.47\textwidth]{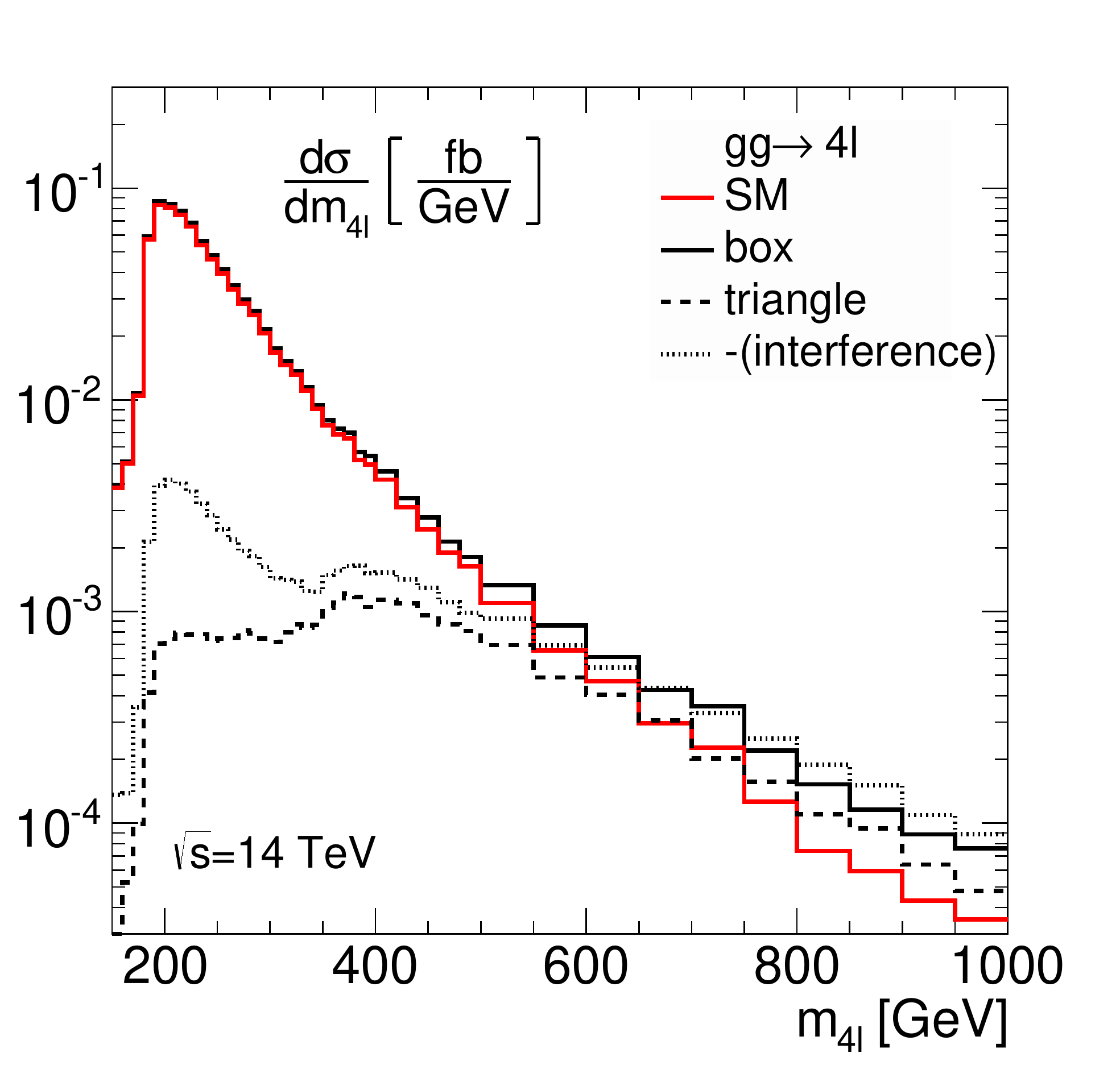}
    \includegraphics[width=.47\textwidth]{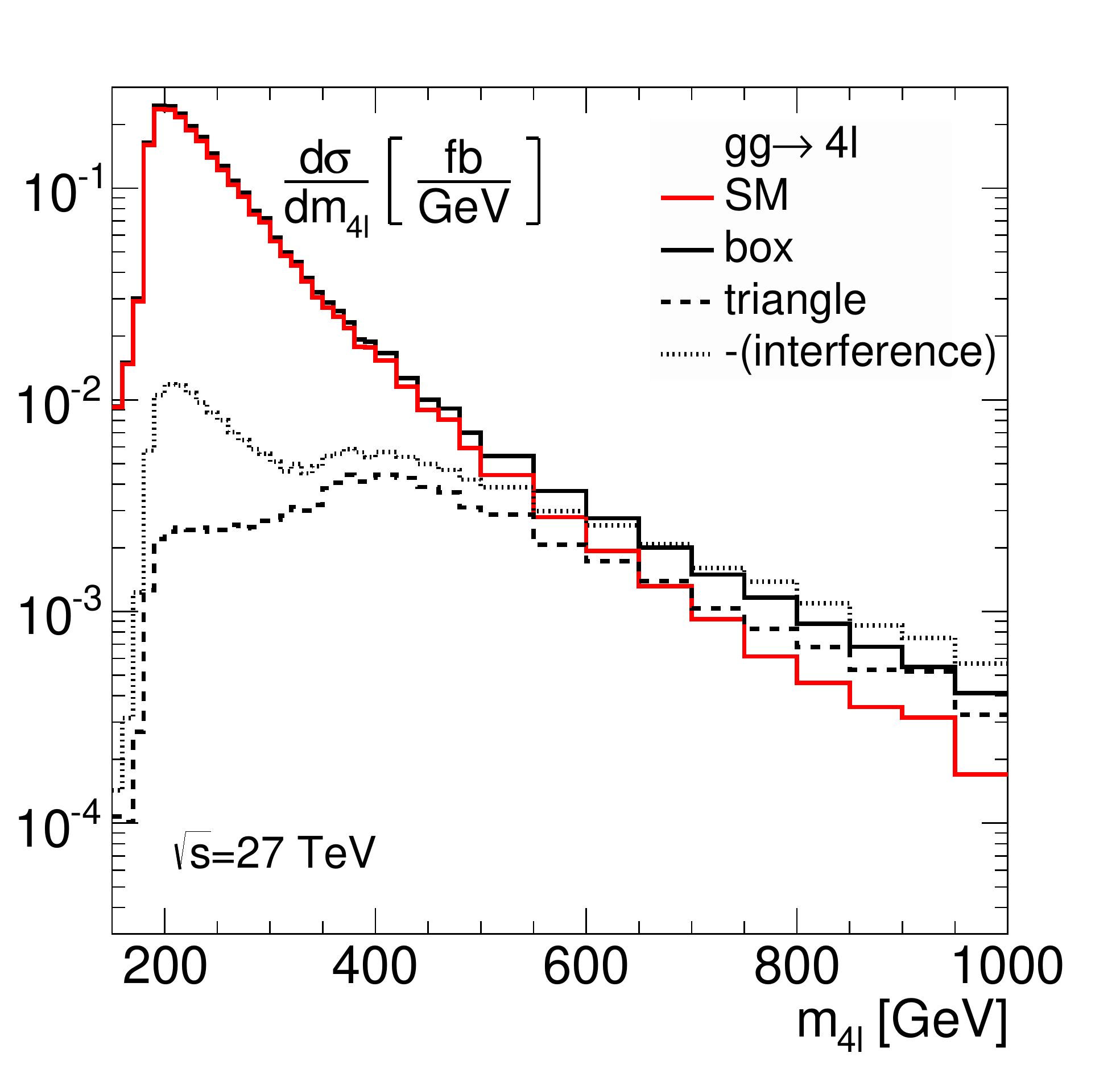}
 \caption{Four-lepton invariant mass distribution for the $gg\rightarrow 4\ell$ process 
 in the full SM (red), triangle component $\sigma_{tt}$ (black dashed), box component $\sigma_{cc}$ (black solid), and the interference between them $\sigma_{tc}$ (black dotted), for the LHC at 14~TeV (left) and 27~TeV (right).}
 \label{fig:mzz}
\end{figure}

It is illustrative to separate the contributions to the gluon fusion production of $Z$ boson pair as
\begin{alignat}{5}
  \frac{d\sigma}{dm_{4\ell}}& =   \frac{d\sigma_{tt}}{dm_{4\ell}}+  \frac{d\sigma_{tc}}{dm_{4\ell}}+  \frac{d\sigma_{cc}}{dm_{4\ell}} \;,
  \label{eq:cutflowm4l} 
\end{alignat}
where $\sigma_{tt}$ corresponds to the Higgs signal contribution, $\sigma_{tc}$ to the signal and box diagram interference, and  $\sigma_{cc}$ to only the box contribution. We show in Fig.~\ref{fig:mzz} the  full $m_{4\ell}$  distribution in the SM, and also individually for each of its components. Remarkably, the $gg\to ZZ$ process 
displays a substantial destructive interference that is larger in magnitude than the contribution from the Higgs signal diagram alone, for the full off-shell $m_{4\ell}$ spectrum. This feature is important in understanding the subsequent results  in the new physics scenarios.

It was pointed out in \cite{offshell} that off-shell Higgs production can be utilized to determine the Higgs boson total width -- a method already adopted by the ATLAS and CMS collaborations~\cite{offshell_ex}. This process is also sensitive to a new color singlet with couplings to top quarks and $Z$ bosons, thus appearing as a new resonance in the $m_{ZZ}$ profile. Additionally, it can probe new colored particles with couplings to the Higgs boson, resolving the long- and short-distance Higgs-gluon interactions~\cite{offshell_model}. The latter feature results in bounds on the top-Higgs Yukawa coupling which are complementary to those from $pp\rightarrow t\bar{t}h$~\cite{legacy}. 
Although there are several final states for the Higgs decay that can be examined, it has been observed that the $ZZ$ final state is optimal $-$ it not only leads to a large interference with the continuum $ZZ$ process above $M_{ZZ}>2M_Z$ as discussed earlier, but also gives rise to a clean four-lepton final state, thereby reducing the experimental systematics on the background estimate~\cite{width}.

We now briefly describe our LHC analysis framework adopted in the subsequent sections for studying the $p p \rightarrow ZZ$ process in the SM and different BSM scenarios. We consider the gluon fusion production of the Higgs boson through heavy quark loops, $gg\to h^*\to ZZ$, and the associated two major backgrounds  processes
\begin{equation}
q\bar{q}\rightarrow ZZ\quad {\rm and}\quad {gg\rightarrow ZZ}.
\end{equation}
The first background arises at the tree level, dominating the event yield, while the second contribution leads to crucial interference effects with the Higgs signal in the off-shell regime.
We have generated the signal and background events using \textsc{MCFM}~\cite{mcfm}, including spin correlations and off-shell effects, particularly for $Z$-decays to lepton pairs. QCD corrections to the gluon fusion subprocess have been incorporated with an overall K-factor~\cite{offshell}. 

We consider the following two setups for the LHC
\begin{eqnarray}
&{\rm HL-LHC:} \quad &14\tev,\quad 3~{\rm ab}^{-1},\\
&{\rm HE-LHC:} \quad &27\tev,\quad 15~{\rm ab}^{-1}.
\end{eqnarray}
For estimating the LHC sensitivity, we have adopted the CMS analysis~\cite{offshell_ex} strategy for favorable signal selection and background suppression, with the kinematical acceptance criteria being as follows
\begin{alignat}{5}
& p_{T\ell}>10~\gev \;,    & |\eta_\ell| &<2.5 \;,  \notag \\
& m_{4\ell}>150~\gev \;, & m_{\ell \ell'} &>4~\gev \;, \notag \\
& m_{\ell \ell}^{(1)} = [40,120]~\gev \;,  \qquad & m_{\ell \ell}^{(2)} &= [12,120]~\gev \;,
\label{eq:cutflowm4l} 
\end{alignat}
where the last two $m_{\ell \ell}$ refer to the leading and sub-leading opposite charge flavour-matched lepton pair. We also demand the leptons to be isolated by requiring $\Delta R_{\ell \ell}>0.2$. We have employed the  \textsc{CTEQ6L1}~\cite{cteq} PDF set and the factorization and renormalization scales are chosen as $\mu_F=\mu_R=m_{4\ell}/2$. The cross-section for the process $gg\to 4\ell$ ($q\bar q\to 4\ell$) increases from 6.1 fb (18 fb) to 19 fb (35 fb) for $m_{4\ell}>200$~GeV with the LHC energy upgrade from 14~TeV to 27~TeV. With this $m_{4\ell}$ requirement, we see that the $gg\to 4\ell$ cross section is increased by about a factor of three.

\section{Virtual Effects from Higgs Portal}
\label{sec:portal}

As a first example, to illustrate the idea that new states responsible for partially addressing the naturalness problem of the Higgs mass can be probed by studying Higgs processes at higher energy scales, we consider a scalar portal to the Higgs sector. The study of a scalar portal to the Higgs sector also has strong motivations in dark matter physics, and in electroweak baryogenesis. A stable scalar singlet particle coupled to the SM sector through the Higgs boson can make up the DM relic density through thermal freeze-out~\cite{inv_portal,cline}. In models of electroweak baryogenesis, in order to achieve a strongly first order phase transition, often new scalars strongly coupled to the SM Higgs doublet are included~\cite{curtin}.

\begin{figure}[t!]
\centering
\vspace{0.6cm}
  \includegraphics[width=.24\textwidth]{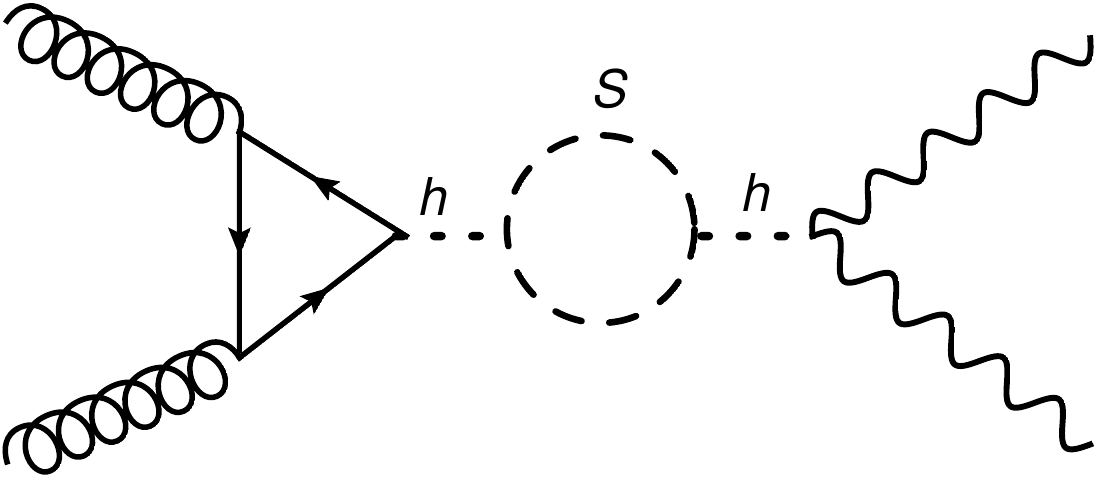}\hspace{0.08cm}
  \includegraphics[width=.24\textwidth]{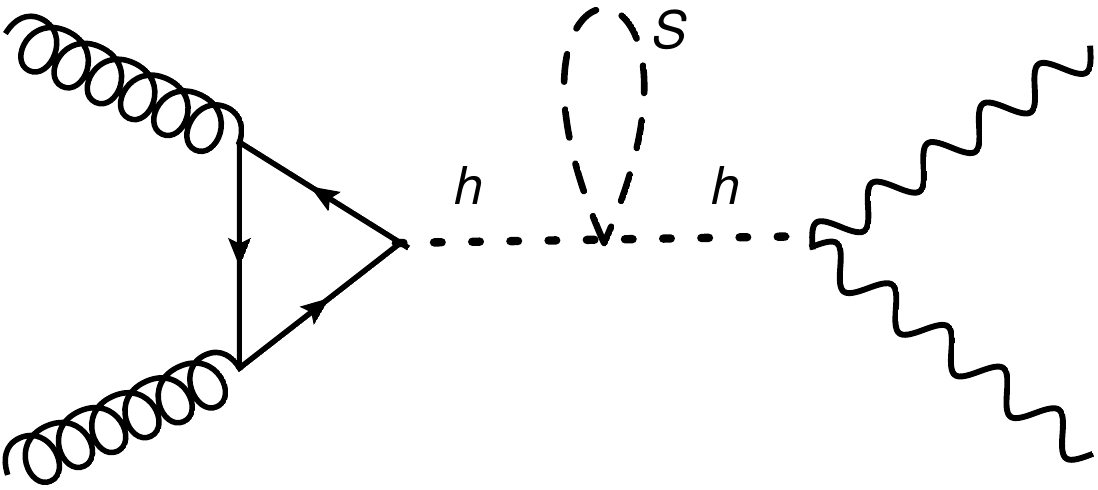}\hspace{0.08cm}
  \includegraphics[width=.24\textwidth]{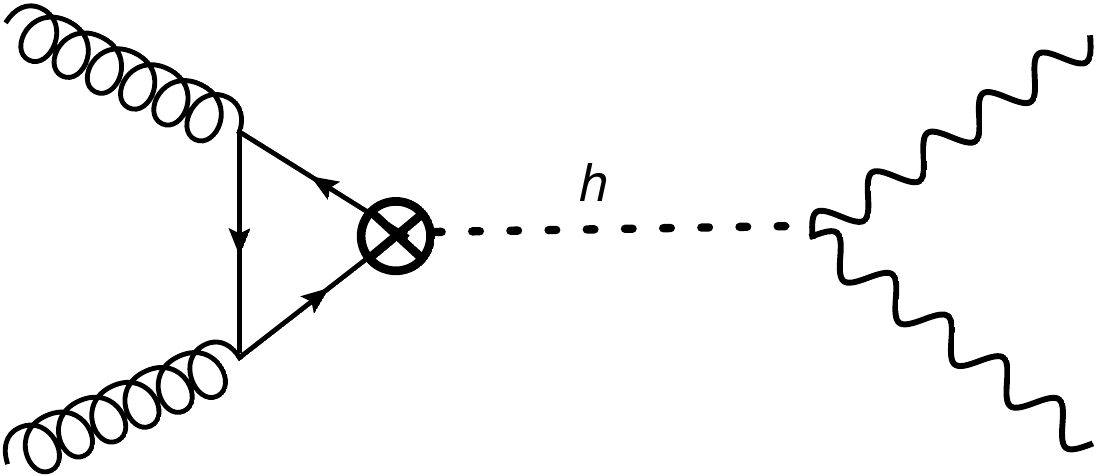}\hspace{0.08cm}
  \includegraphics[width=.24\textwidth]{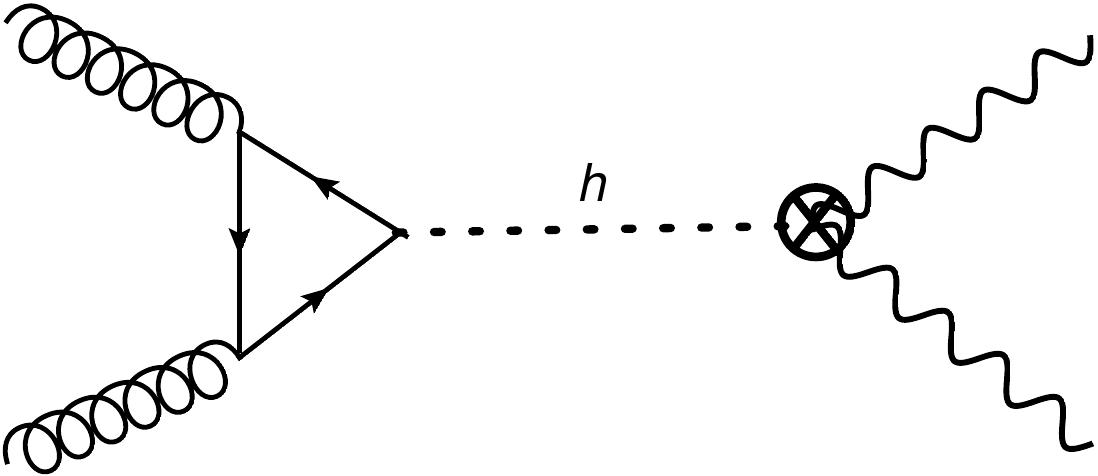}
 \caption{Representative set of Feynman diagrams for the one-loop corrections  to $gg\rightarrow ZZ$ production from the singlet scalar sector.}
 \label{fig:feyndiag2}
\end{figure}

We have discussed such a scenario in a recent study~\cite{Goncalves:2017iub}. 
The scenario can be described by a simple low-energy effective Lagrangian as follows
\begin{equation}
\mathcal{L} \supset \partial_\mu S \partial^\mu S^* - \mu^2|S|^2 - \lambda_S |S|^2 |H|^2,
\label{eq:lag}
\end{equation}
where $S$ is a complex singlet scalar field odd under a $\mathcal{Z}_2$ symmetry, with the SM fields being even under it. After electroweak symmetry breaking, the mass of the singlet is given by $m_S^2 = \mu^2+\lambda_S v^2/2$, where $v=246$ GeV is the vacuum expectation value of the Higgs field.

A new scalar $S$ with couplings to the Higgs field has implications for the naturalness problem. In the presence of new states of mass-scale $\Lambda$, directly or indirectly coupled to the Higgs boson, all quadratic sensitivity of the Higgs mass to the scale $\Lambda$ should be cancelled in a natural theory. To begin with, the leading one-loop correction to the high-scale Higgs mass $M_h$, from the top quark and the scalar singlet loops is given as 
\begin{eqnarray}
\delta M_h^2  &=& \frac{1}{16 \pi^2} (\lambda_S - 2 N_c y_t^2) \Lambda^2 + \frac{6 N_c y_t^2}{16 \pi^2}  m_t^2 \log \frac{\Lambda^2}{m_t^2}  \nonumber \\
                      & - &\frac{1}{16 \pi^2} \left(\lambda_S m_S^2+\lambda_S^2 v^2\right)\log \frac{\Lambda^2}{m_S^2} ,
\end{eqnarray} 
where $y_t$ is the top quark Yukawa coupling in the SM and the number of colours $N_c=3$. If we impose the high-scale parameter relation $\lambda_S(\Lambda^2) = 6 y_t^2(\Lambda^2)$, the quadratic divergent contribution to the Higgs boson mass from the top quark loop is cancelled by the opposite-sign contribution from the scalar singlet loop. In a UV-complete theory, such a relation can ensue from an underlying symmetry, for example, in a supersymmetric theory the scalar top loops cancel the top quark loop contributions. Partners of the top quark that do not possess SM color or electroweak charges can also arise from different class of symmetries protecting the Higgs mass, as in the neutral naturalness scenarios~\cite{Twin,Folded_SUSY,Orbifold}. 
 One of the simplest realizations of this idea is the twin Higgs model~\cite{Twin}, which can be generalized to a broader class of supersymmetric~\cite{Folded_SUSY} or non-supersymmetric orbifold Higgs models~\cite{Orbifold}. Such neutral naturalness scenarios also predict strikingly different signatures at collider experiments~\cite{signals_twin}.


\begin{figure}[t!]
\centering
  \includegraphics[width=0.48\textwidth]{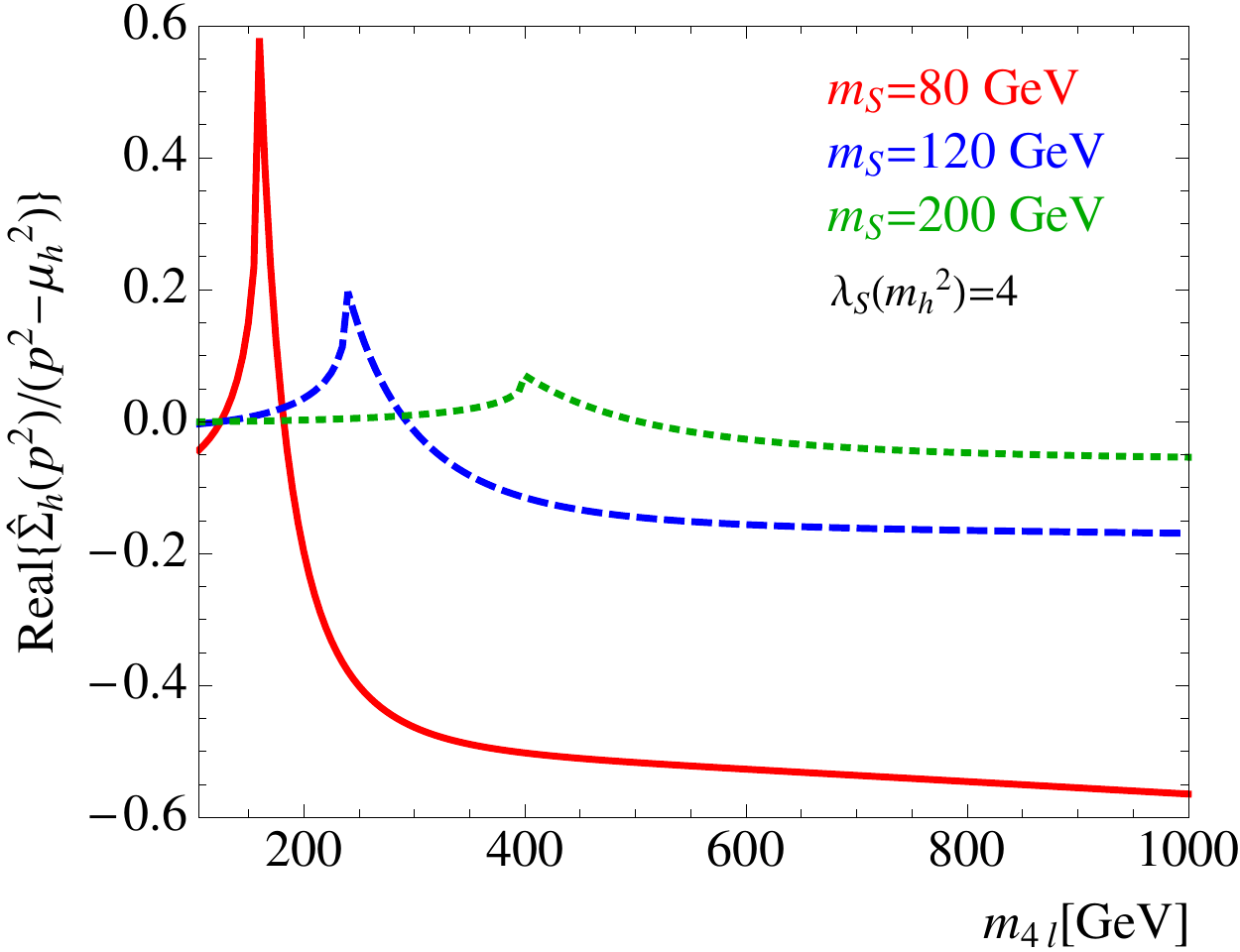}
  \includegraphics[width=0.47\textwidth]{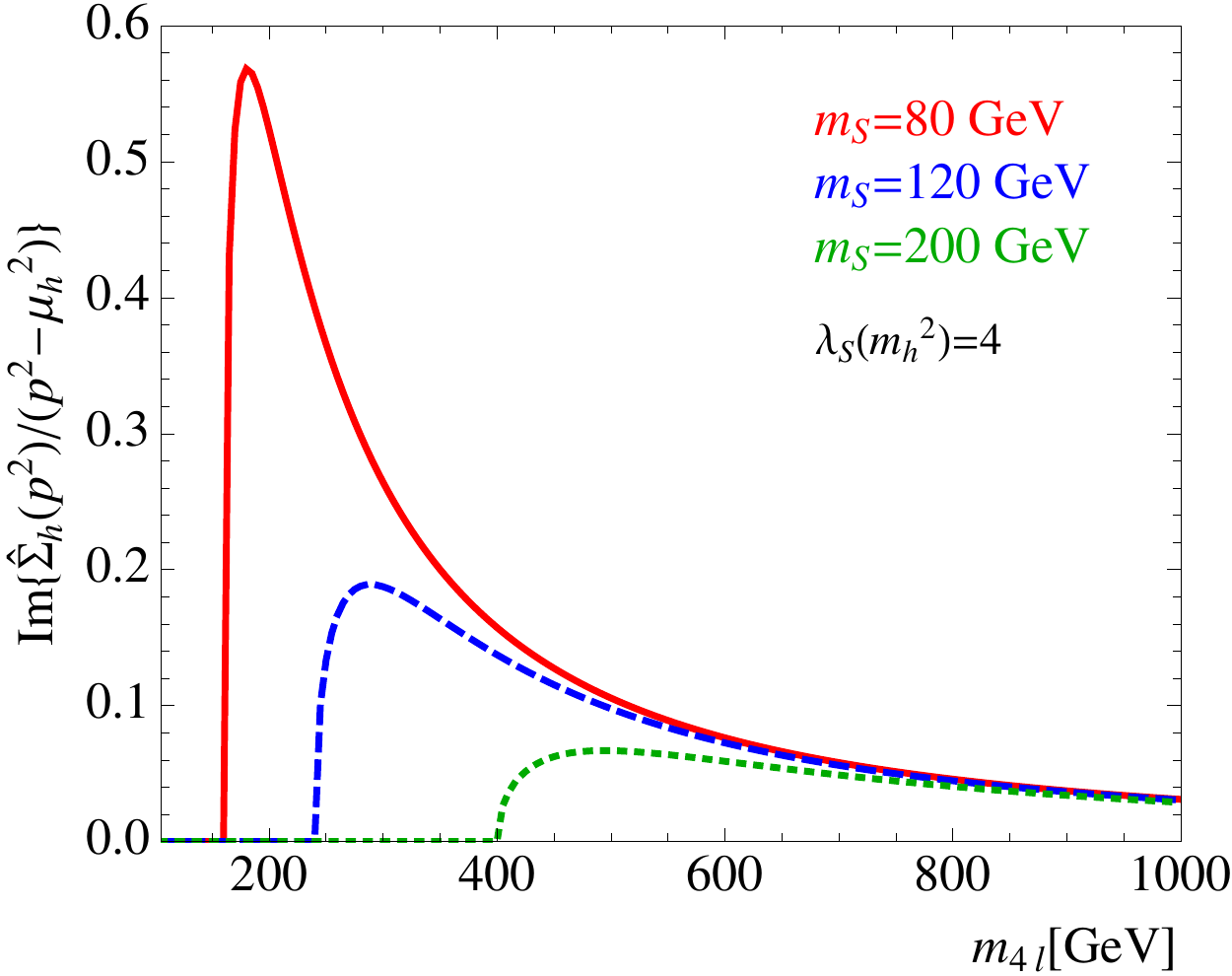}
 \caption{Real (left) and imaginary (right) parts of the Higgs boson
 renormalized self-energy corrections $\hat{\Sigma}_h$, scaled by the propagator factor $p^2-\mu_h^2$, as a function of $m_{4\ell}$.}
 \label{fig:selfE}
\end{figure}

For probing the existence of such a maximally hidden scalar sector, the key observation is that the singlet scalar sector would lead to NLO electroweak corrections to the process $gg\rightarrow ZZ$, representative Feynman diagrams for which are shown in Fig.~\ref{fig:feyndiag2}. These corrections constitute a separately renormalizable, gauge-invariant,  UV-finite subset. 

In our computation of the electroweak radiative corrections, we follow the {\it complex mass scheme}~\cite{complex_mass}, in which  the renormalized Higgs boson self-energy is defined as 
\begin{alignat}{5}
& \hat{\Sigma}_{h}(p^2)=\Sigma_{h}(p^2)-\delta\mu^2_h+(p^2-\mu^2_h)\delta Z_h\;,
\label{eq:complexmass1} 
\end{alignat}
where the complex Higgs mass squared is ${\mu_h^2=m_h^2-im_h\Gamma_h}$ and the renormalization constants are defined as
\begin{equation}
 \delta \mu_h^2=\Sigma_{h}(\mu_h^2)\;, \qquad \delta Z_h=-\frac{d\Sigma_{h}}{dp^2} (\mu_h^2).
\label{eq:complexmass2} 
\end{equation}
Throughout our analysis, we have evolved $\lambda_S(Q^2)$ using the renormalization group equation at one-loop.

We now briefly discuss the qualitative features of the one-loop contributions from the scalar singlet sector. We show the behaviour of the Higgs boson self-energy corrections $\hat{\Sigma}_h$ (scaled by the propagator factor $p^2-\mu_h^2$) as a function of the sub-process centre of mass energy $m_{4\ell}$ in Fig.~\ref{fig:selfE}. While there is a resonant enhancement in the real part of the self-energy correction, the imaginary part shows a threshold behaviour near the energy scale ${m_{4\ell}=2m_S}$. As we shall see in the following, these features lead to interesting consequences in the differential distributions for the LHC processes that we study next.
 In order to determine the effect of these electroweak corrections, we propose to study the ${pp\rightarrow Z^{(*)}Z^{(*)} \rightarrow 4\ell}$ channel at the LHC, the framework for which is discussed in Sec.~\ref{sec:process}. 

\begin{figure}[t]
\centering
  \includegraphics[width=0.335\textwidth]{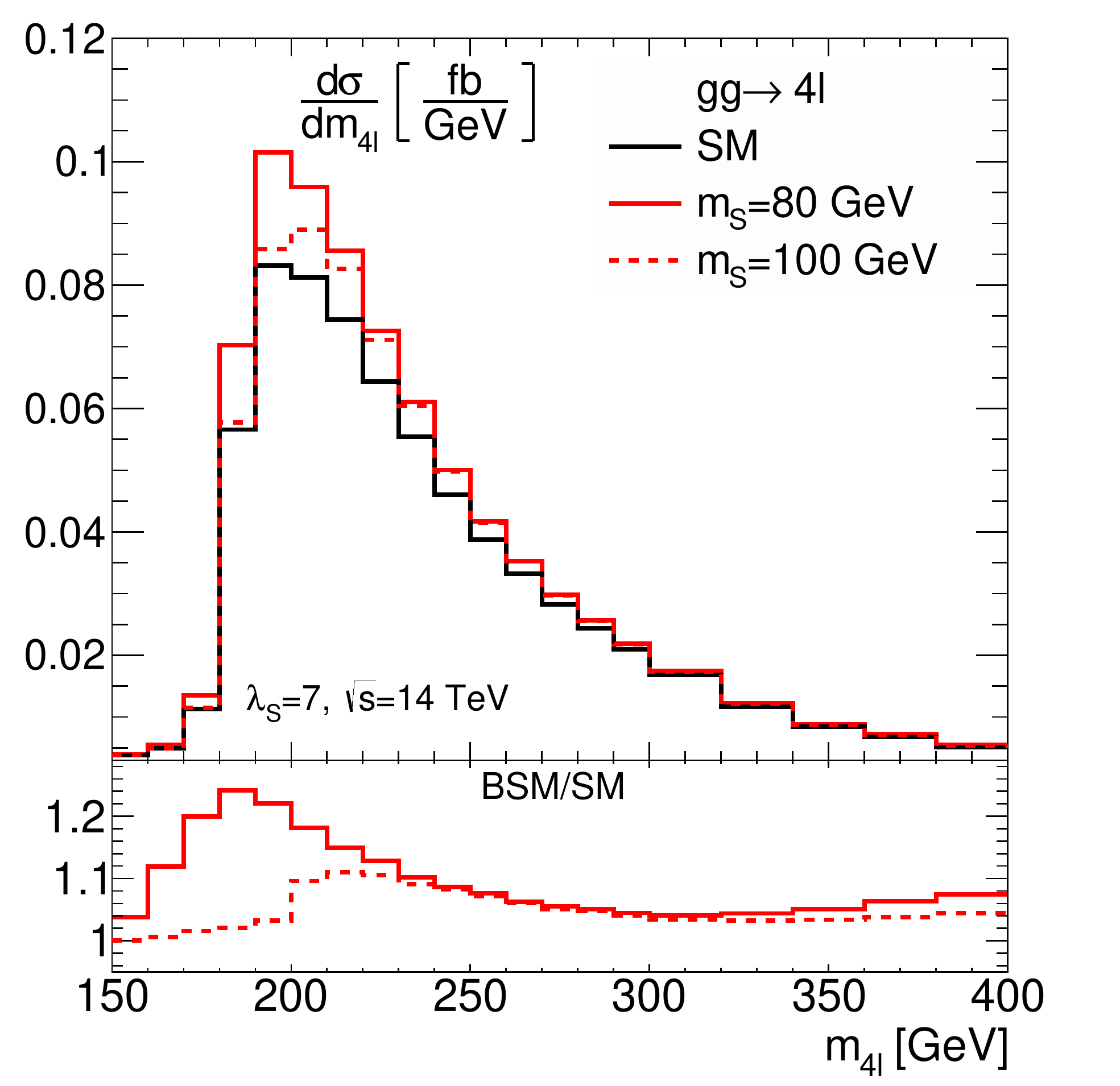}\hspace{-0.45cm}
    \includegraphics[width=0.335\textwidth]{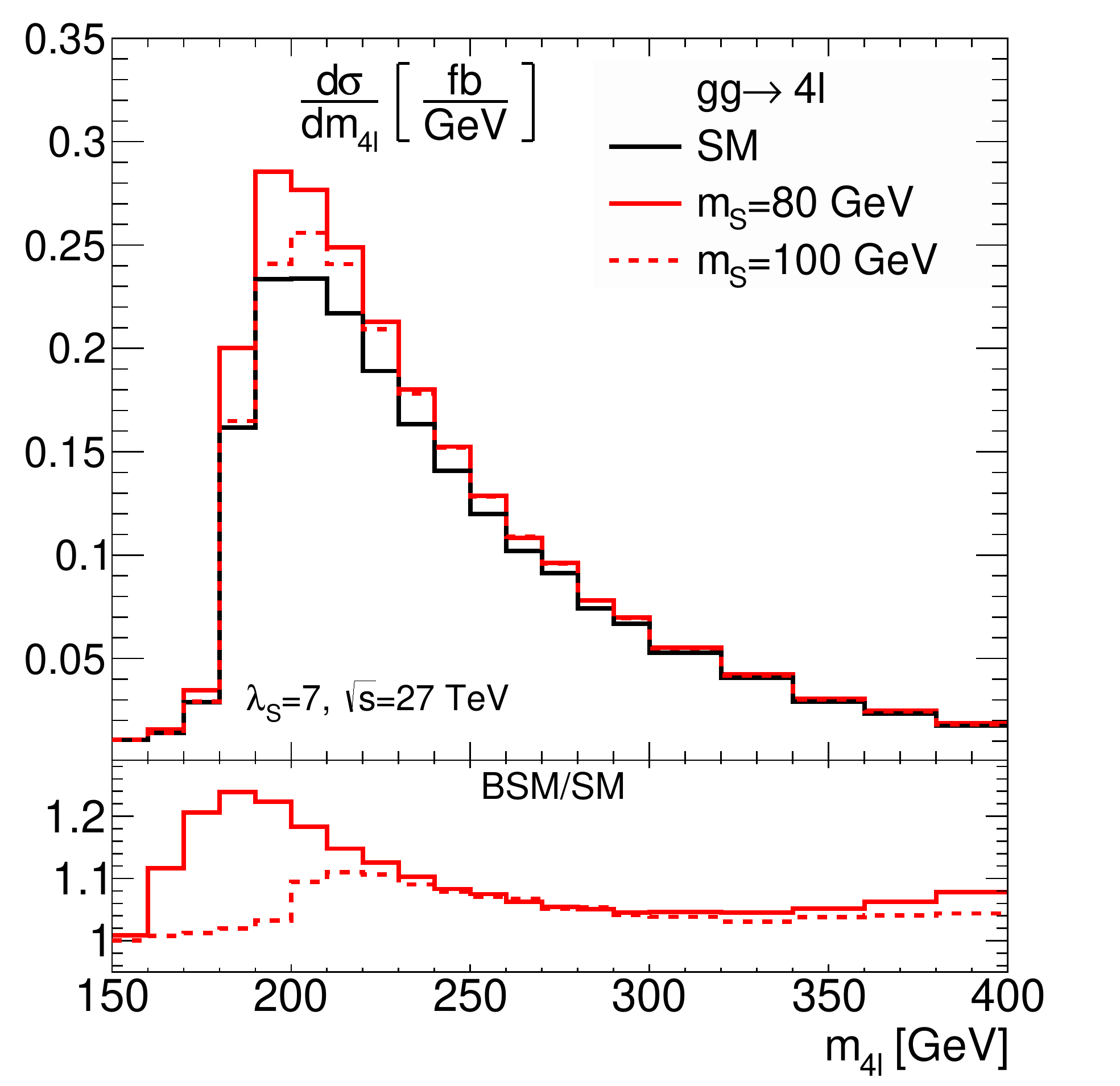}\hspace{-0.45cm}
    \includegraphics[width=0.37\textwidth]{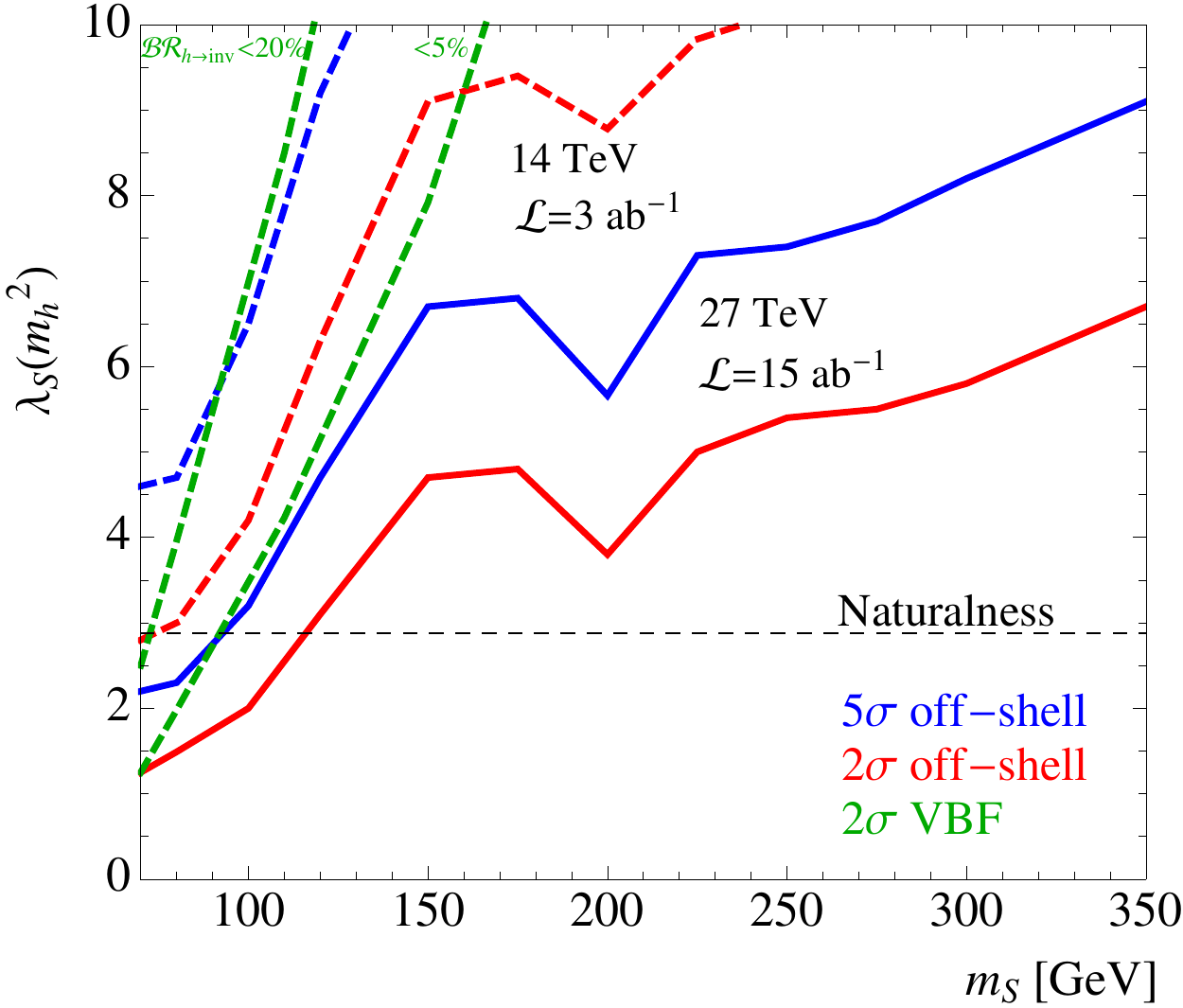}
 \caption{Four-lepton invariant mass distribution for the $gg\rightarrow 4\ell$ process at the LHC $14$~TeV (left) and 27 TeV (center) in the SM (black) and in the presence of an additional gauge singlet scalar (red), including the one-loop EW effects from the singlet scalar sector. 
 We show the signal ratio between the scalar singlet model and the SM in the bottom panels. 
 Right: $2\sigma$ (red) and $5\sigma$ (blue) sensitivity on the singlet-Higgs coupling $\lambda_S$ at the scale $m_h^2$  as a function of the singlet scalar mass $m_S$  from the off-shell Higgs analysis at the 14~TeV LHC
with $\mathcal{L}=3$~ab$^{-1}$  (dashed) and at the 27~TeV LHC, with $\mathcal{L}=15$~ab$^{-1}$ (solid).
For comparison, we also show the reach from 
the weak-boson fusion production of Higgs above its threshold, assuming the high-luminosity LHC $2\sigma$ confidence level projections of ${\mathcal{BR}(h\rightarrow \text{invisible})<20\%}$  (green dotted) and  $5\%$  (green dashed). }
 \label{fig:m4l}
\end{figure}

In Fig.~\ref{fig:m4l} (left and center), we present the four-lepton invariant mass distribution at the LHC for the $gg\rightarrow 4\ell$ process in the SM (black solid line) and in the model with an additional scalar gauge singlet (red solid and dashed lines), for different choices of the mass of the scalar singlet $m_S$. We see that in addition to shifting the on-shell Higgs rate~\cite{Han, Englert:2013tya}, the higher order corrections to $gg\rightarrow 4\ell$ in the singlet model also result in relevant  kinematic features in the $m_{4\ell}$ distribution, especially above the threshold $m_{4\ell}>2m_S$. We show the signal ratio between the scalar singlet model and the SM in the bottom panels, and find that for $m_S=80$~GeV and $\lambda_S(m_h^2)=7$, the SM predictions could be modified by up to $25\%$ near the $2m_S$ threshold.

To estimate the sensitivity at the LHC for the singlet sector parameter space $(m_S,\lambda_S)$,
we perform a binned log-likelihood analysis based on the CL$_{s}$ method, using the $m_{4\ell}$ distribution~\cite{read}. The results are presented in Fig.~\ref{fig:m4l} (right) with the $2\sigma$  and $5\sigma$ sensitivity on $\lambda_S$ (evaluated at the scale $m_h^2$) shown as a function of the singlet scalar mass $m_S$. The black-dashed line shows the value of $\lambda_S(m_h^2)$ for which the high-scale parameter relation $\lambda_S(\Lambda^2) = 6 y_t^2(\Lambda^2)$ is satisfied at $\Lambda=10$ TeV, where the latter choice is motivated to address the little-hierarchy problem~\cite{Barbieri}. The coupling values at different scales have been related by the renormalization group evolution. We see that the HE-LHC upgrade can access singlet scalar masses of around $120$ GeV at the $2\sigma$ confidence level, for couplings implied by the naturalness relation. It is observed that there is an enhancement of sensitivity of the off-shell channel for values of $m_S$ close to $m_t$. This is because of the opening of two different thresholds close to each other, namely, the $2m_t$ threshold in the triangle and box diagrams for $ZZ^*$ production, and the $2m_S$ threshold in the radiative correction from the scalar singlet to the same process.

The scalar singlets can also be directly pair-produced using the vector-boson fusion (VBF) process with the Higgs produced above threshold, and looked for in the jets and missing momentum final state~\cite{WBF}, namely, $q\bar{q} \rightarrow q \bar{q} h^{(*)} \rightarrow q \bar{q} (S^*S)$, where the $S$ particles are on-shell. For comparison with the off-shell Higgs analysis described above, we present an estimate of the $2\sigma$ reach for the VBF channel as well. To this end, we translate the projected upper bound on the invisible branching ratio for on-shell Higgs in the VBF process, 
$\sigma_{VBF} (h) \mathcal{BR} (h \rightarrow {\rm invisible}) = \sigma_{\rm VBF} (h^{(*)}\rightarrow S^*S)$.
In Fig.~\ref{fig:m4l}, we have shown the reach in the VBF mode assuming two high-luminosity $14$ TeV LHC upper bounds of ${\mathcal{BR}(h\rightarrow \text{invisible})<20\%}$ and $5\%$ at the $2\sigma$ confidence level~\cite{CMS:2017cwx}. The former corresponds to a realistic projection of the systematic uncertainties on the background prediction, while the latter case represents an idealistic limit. We observe that in almost the entire singlet mass range of interest, $m_S>m_h/2$, the off-shell Higgs analysis leads to a better sensitivity on $\lambda_S$ compared to a realistic estimate for the VBF channel.

Since the scalar singlet can serve as a component of the total dark matter (DM) density, it is natural to ask if the constraints from dark matter direct detection in the underground nuclear recoil experiments become relevant or not. Even though the Higgs portal coupling implies a large spin-independent scattering cross-section in these direct detection experiments, the event rate would be small, making such a scenario evade constraints from these searches. This is because a large annihilation rate in the early Universe, which follows from the large coupling with the Higgs required by the naturalness condition, implies a small number density surviving after thermal freeze-out~\cite{cline}. Therefore, the collider probe in off-shell Higgs presented above can become one of the best hopes of detecting such DM particles.

\section{Quantum Criticality}
\label{sec:conformal}

\begin{figure}[b!]
\centering
  \includegraphics[width=0.45\textwidth]{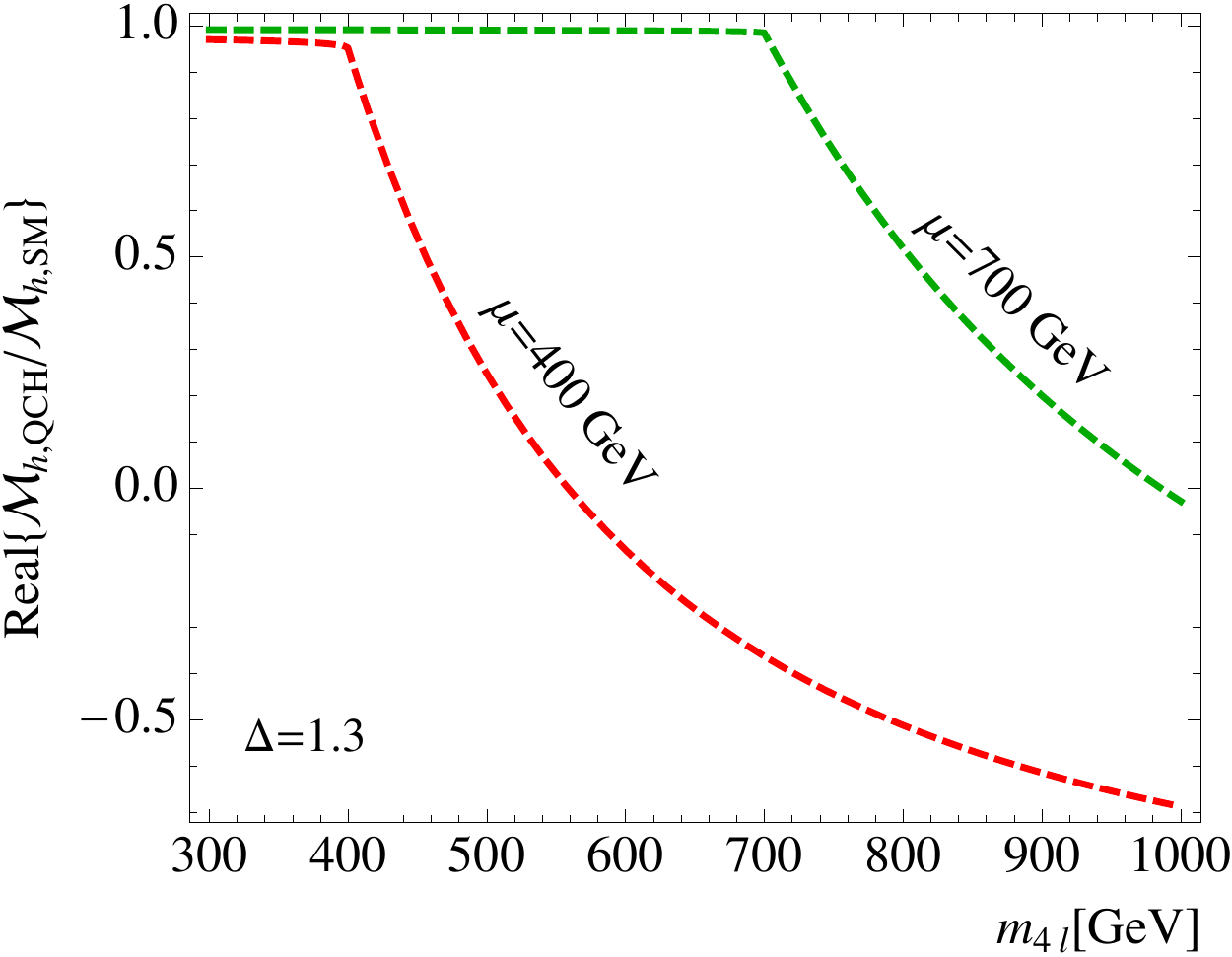}
  \includegraphics[width=0.45\textwidth]{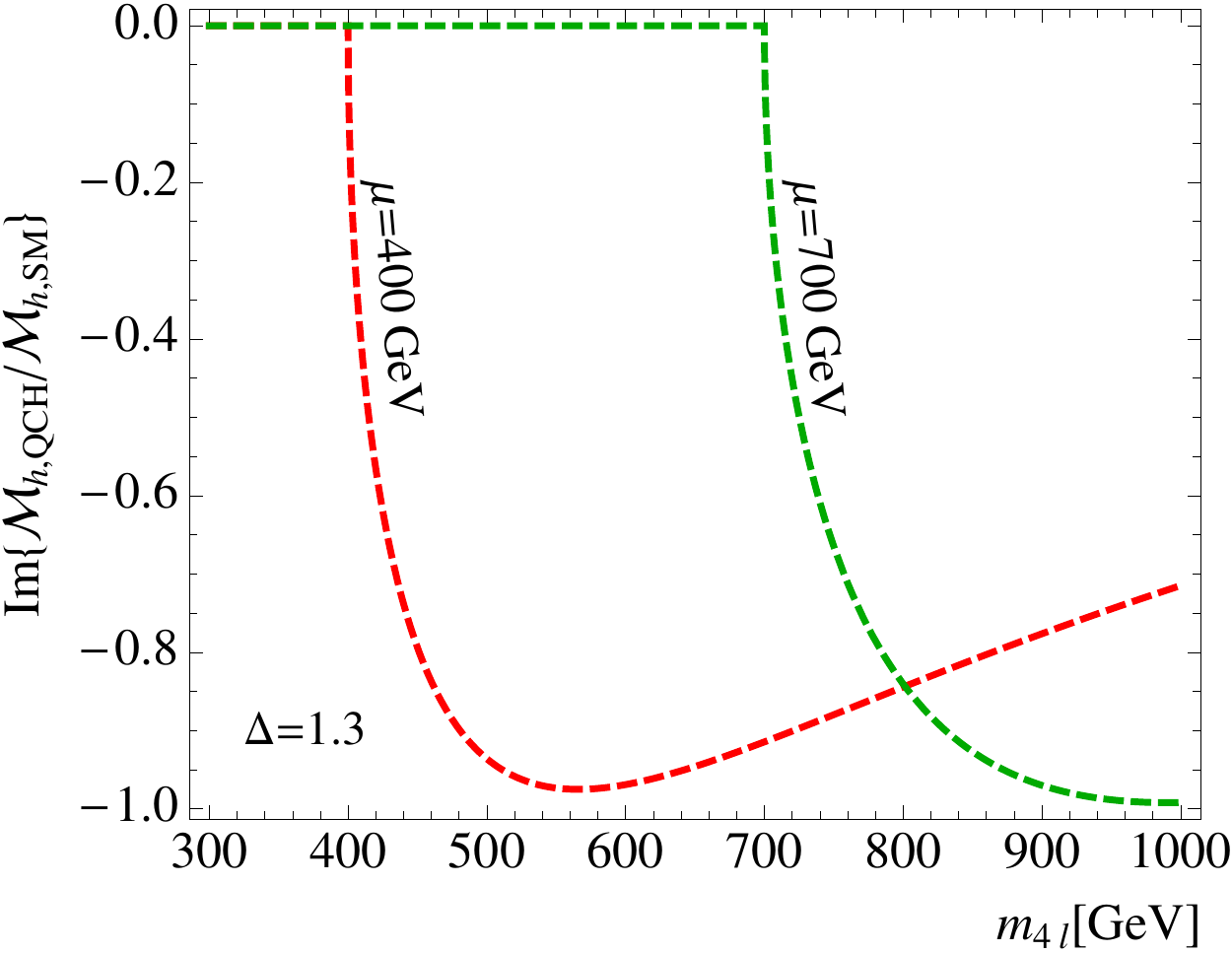}
 \caption{Real (left) and imaginary (right) components for the amplitude ratio between the s-channel Quantum Critical Higgs $\mathcal{M}_{h,QCH}$ and SM $\mathcal{M}_{h,SM}$ as a function of $m_{4\ell}$. We present two BSM scenarios $\mu=400$~GeV (red) and $\mu=700$~GeV 
(green), assuming $\Delta=1.3$.}
 \label{fig:branchcut}
\end{figure}

\begin{figure}[t!]
\centering
\includegraphics[scale=0.256]{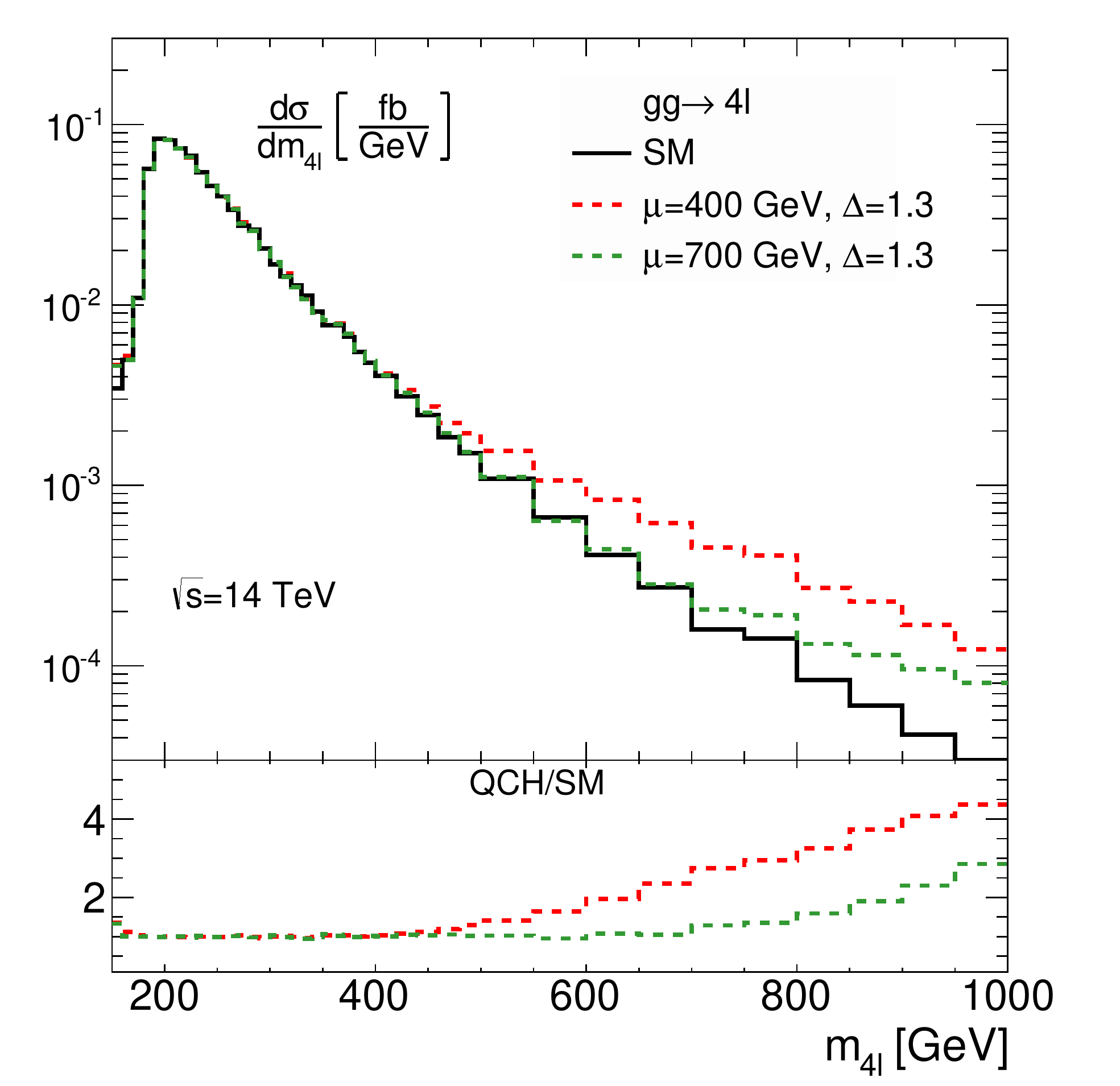}\hspace{-.45cm}
\includegraphics[scale=0.256]{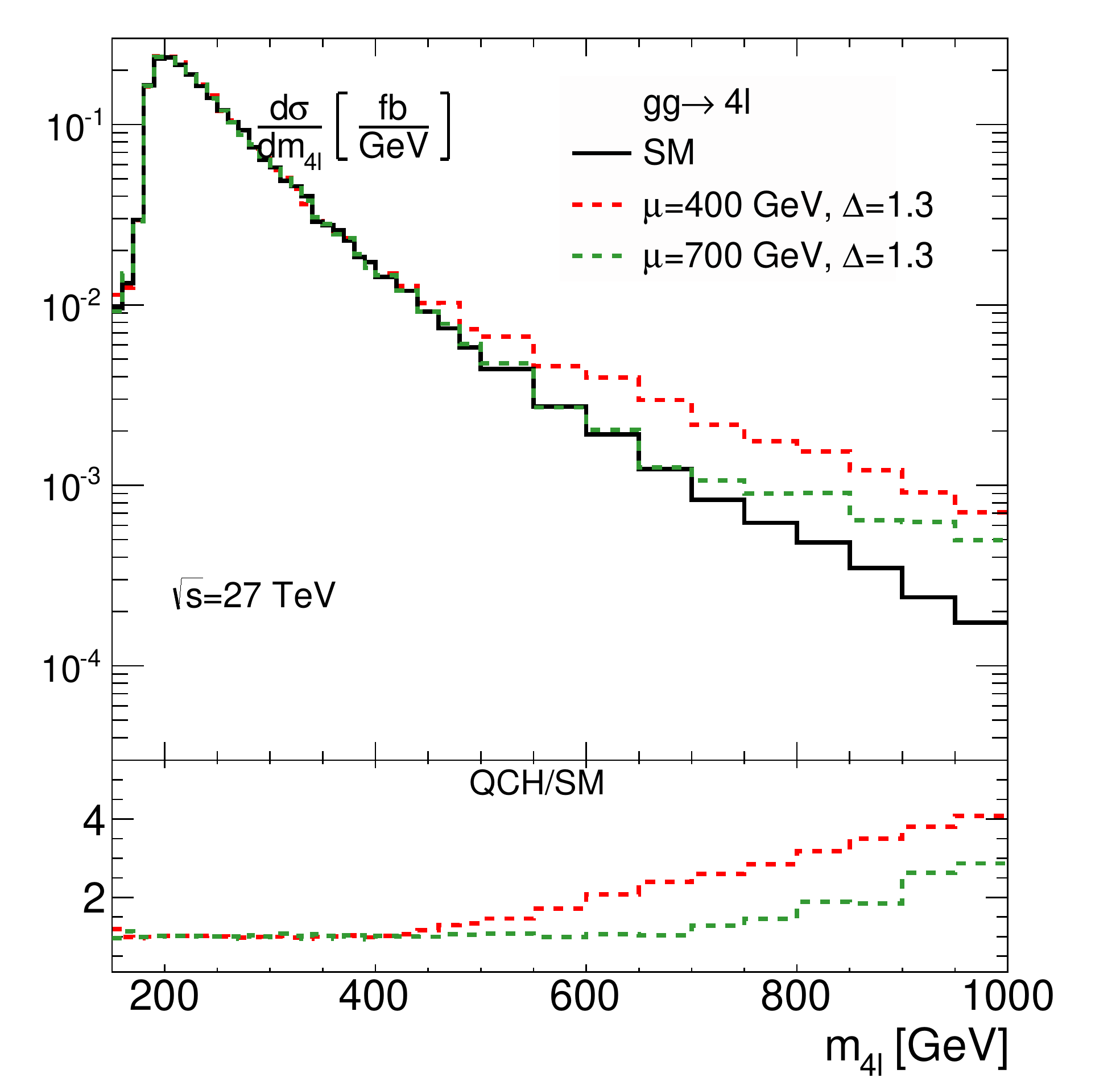}\hspace{-.45cm} 
 \includegraphics[width=0.367\textwidth]{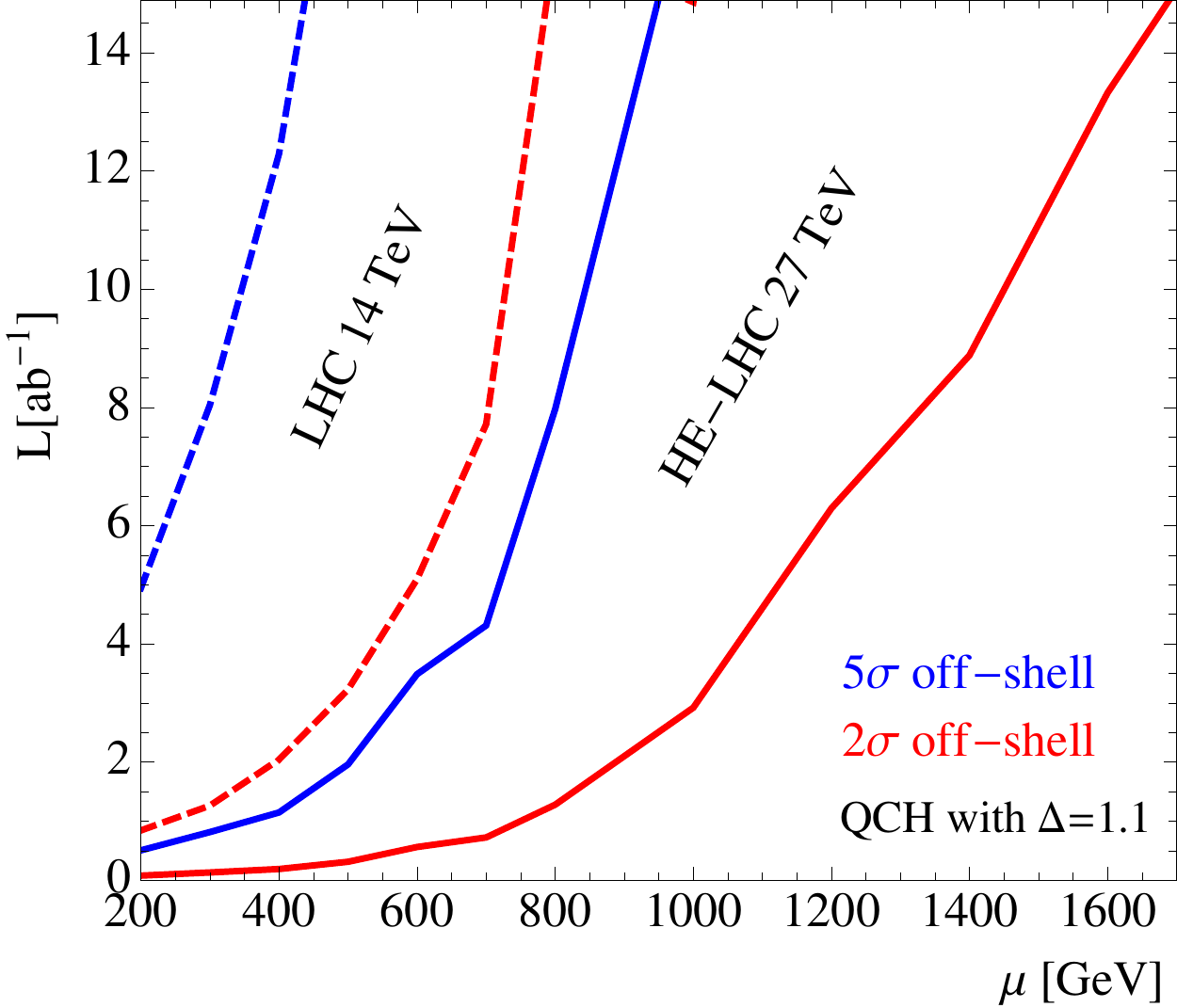}
\caption{Four-lepton invariant mass distribution for the $gg\rightarrow 4\ell$ process at the LHC $14$~TeV (left) and 27 TeV (center) for the SM (black) and Quantum Critical Higgs with $\Delta=1.3$ and $\mu=400$~GeV (red) and $\mu=700$~GeV (green).
We show the signal ratio between the QCH model and the SM in the bottom panels.
Right: $5\sigma$ (blue) and $2\sigma$ (red) bounds on the conformal symmetry breaking scale $\mu$. 
We show results for the 14~TeV LHC (dashed) and the  27~TeV HE-LHC (solid), assuming $\Delta=1.1$. 
}
\label{fig:m4l_QCH}
\end{figure}

Inspired by certain condensed matter systems in which a light scalar excitation could occur by tuning parameters close to a critical value for a second-order phase transition, the authors of Ref.~\cite{Bellazzini:2015cgj} considered a system with an approximate scale invariance at the critical point. If the system presents a non-trivial fixed point, then the non-trivial critical exponents characterized by the scaling dimensions ($\Delta$) imply possible dramatic changes of the field properties and could even lead to non-particle description. Practically, beside the light Higgs boson as an excitation near a critical point, there may be a continuum in the spectrum associated with the dynamics underlying the phase transition at the zero temperature (quantum phase transition), not far from the Higgs resonance. An attractive  consequence is that the quantum corrections to Higgs boson mass would have a weaker dependence on the cutoff scale. For instance, the top-quark loop contribution would be modified as \cite{Bellazzini:2015cgj,Stancato:2008mp}
\begin{equation}
\delta m_h^{4-2\Delta} = {3 \lambda_t^2\over 8\pi^2} \Lambda^{4-2\Delta} .
\label{eq:dmh}
\end{equation}
Thus a scaling dimension larger than the SM value, $\Delta>1$, would alleviate the Higgs mass fine-tuning with respect to the corrections from the higher scale $\Lambda$.
The same underlying dynamics may lead to observable effects on the Higgs couplings as well as the propagation. 
The $ZZh$ coupling,  top-Yukawa coupling,  and  Higgs propagator can be cast into the forms 
\begin{eqnarray}
&  g_{ZZh} &= g^{\mu\nu}\Gamma_{ZZh}, \quad 
y_t=\sqrt{2}\frac{m_t}{v} \left(\frac{\Lambda}{v}\right)^{\Delta-1} \;,  \\
& G_h(p) &= -{iZ_h \over {(\mu^2 - p^2-i\epsilon)^{2-\Delta} - (\mu^2 - m_h^2)^{2-\Delta} }},\quad  
Z_h = {2-\Delta\over (\mu^2 - m_h^2)^{\Delta-1}} ,
\end{eqnarray}
where  $\Gamma_{ZZh}$ is a momentum-dependent form factor, with  scaling dimension $1\le\Delta\le1.5$, IR transition scale $\mu>m_h$,
and $\Lambda$ is the UV cut-off scale~\cite{Stancato:2008mp}. Therefore, the Higgs two-point function is given by a pole  at the  Higgs mass $m_h$ and a branch cut above the conformal symmetry breaking scale $p^2>\mu^2$. The SM predictions can be recovered upon taking the limit $\Delta\rightarrow 1$.

While the on-shell Higgs measurements are largely insensitive to the scale $\mu$~\cite{unhiggs}, the presence of this continuum spectrum for $p^2>\mu^2$ can be probed by the off-shell Higgs measurement. In Fig.~\ref{fig:branchcut} we show the real and imaginary components for the amplitude ratio between the $s$-channel QCH  $\mathcal{M}_{h,QCH}$ and SM $\mathcal{M}_{h,SM}$ as a function of $m_{4\ell}$. 
Although the QCH displays small corrections associated to the real part of the amplitude for $m_{4\ell} < \mu$, the presence of the branch-cut at $m_{4\ell} = \mu$ results into large contributions above the scale $\mu$.
In Fig.~\ref{fig:m4l_QCH}, we show how these corrections translate into the  $m_{4\ell}$ distribution for the $gg\rightarrow 4\ell$ process.  
We find significantly large effects at the LHC, see the left and center panels. We show the signal ratio between the QCH model and the SM in the bottom panels and we see that the ratio could be as high as a factor of $3-4$ at the higher invariant mass region.
We can probe $\mu\sim 500$~GeV for a $2\sigma$ exclusion at the HL-LHC, and 
$\mu\sim 900$~GeV for a $5\sigma$ observation at the HE-LHC, assuming $\Delta=1.1$.

\section{Weakly Coupled Scenario: RG Evolution}
\label{sec:RGE}

The energy-scale dependence of coupling and mass parameters is a fundamental prediction in quantum field theory. The specific form of this running depends on the particle spectrum and their interactions in the underlying theory. The best example thus far is the running of the strong interaction coupling strength ($\alpha_S$), that
has been experimentally probed over a broad energy range, being in excellent agreement with the SM prediction of asymptotic freedom. Including data ranging from tau-decays, deep-inelastic scattering, decay of heavy quarkonia, measurement of jet shapes at $e^+e^-$ colliders, electroweak precision fits, to the present day hadron collider data from the Tevatron and the LHC, the value of $\alpha_S$  has been determined in the energy range of around $2$~GeV to more than $1000$~GeV~\cite{PDG}. Such measurements are not only crucial to test the SM predictions across many orders of magnitude in energy scale~\cite{PDG}, they also furnish some of the most model-independent bounds on new states with color charge running in the loop, independent of their decay properties~\cite{Schwartz,Michael}. It has also been suggested that the determination of the scale-dependence of electroweak gauge couplings using future precision measurements of the Drell-Yan process at high-energy hadron colliders can probe the presence of new particles charged under the SM electroweak interactions~\cite{Ruderman}.

\begin{figure}[t]
\centering
\includegraphics[scale=0.53]{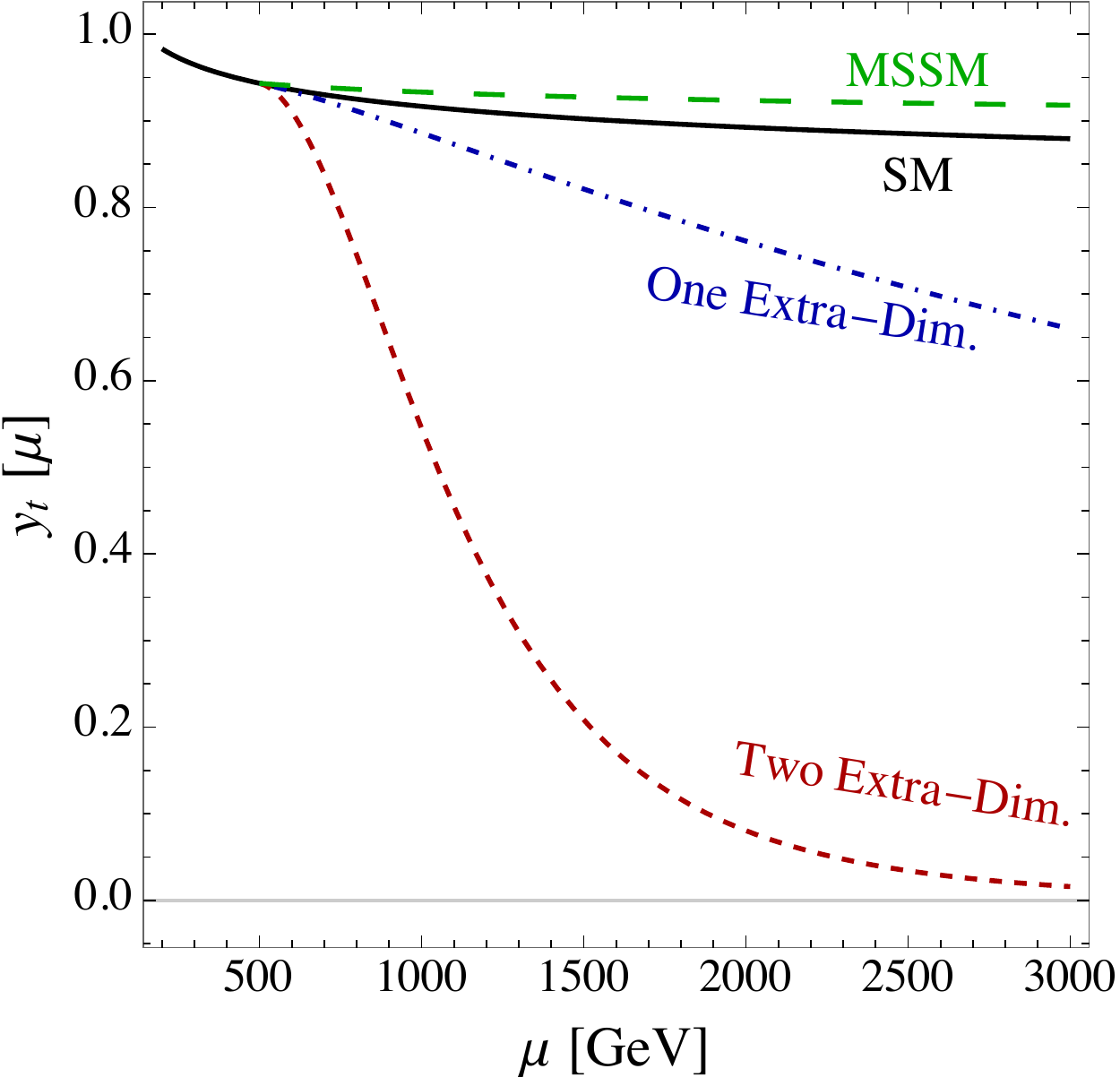}
\caption{LO RGE running of top Yukawa $y_t$ as a function of the energy scale $\mu$, in the SM (black solid), in MSSM (green long-dashed), a model with one extra dimension (blue dot-dashed) and two extra dimensions (red short-dashed). In the MSSM case the common mass of the sparticles is taken to be $500$ GeV. In the extra-dimensional scenarios (with inverse radius $1/R = 500$ GeV) only the SM gauge fields are assumed to propagate in the bulk, while the matter fields are confined to the brane. 
}
\label{fig:yt_SM}
\end{figure}

Studying the energy scale dependence of the Higgs couplings under the renormalization group evolution can also hold clues to new states coupled to the Higgs sector in particular, and the SM particles in general. A first target would be the Higgs coupling to the top quark. Let us begin with a review of the SM Yukawa coupling and then go on to discuss different weakly-coupled beyond SM extensions. In the SM, the dominant contribution to the RG running of the top Yukawa is from QCD corrections, and a sub-dominant but important contribution stems from the top Yukawa itself. There are two reasons for the latter contribution to be important: $y_t$ itself is $\mathcal{O}(1)$ at the scale $\mu = m_h$, and the sign of its contribution to $\beta_{y_t}$ is positive, in contrast to the sign of the gauge contributions, which are negative. At leading order (LO), the RG evolution of $y_t$ is given in the $\overline{\rm MS}$ scheme by 
\begin{equation}
\frac{dy_t}{d t} = \beta_{y_t}^{\rm SM} =\frac{y_t}{16 \pi^2} \left (\frac{9}{2}y_t^2 - 8 g_3^2 - \frac{9}{4} g_2^2 - \frac{17}{20} g_1^2 \right), 
\end{equation}
with $t=\ln(\mu)$. The SM gauge couplings evolve with the energy scales as
\begin{equation}
\frac{dg_i}{dt} = \frac{b_i g_i^3}{16 \pi^2}, 
\end{equation}
at one-loop, with the coefficients  $b_i$ for the gauge couplings $(g_1,g_2,g_3)$ given as
\begin{align}
b_i^{\rm SM} = & (41/10,-19/6,-7) .
\end{align}
We show the LO RGE running of top Yukawa $y_t$ as a function of the energy scale $\mu$ in the SM in Fig.~\ref{fig:yt_SM} (black solid curve). In the energy scales accessible in near future colliders, the change in $y_t$ is observed to be rather small, for example,  $y_t(\mu=5 {~\rm TeV})$ is found to be around $14\%$ smaller compared to $y_t(m_h)$. As we shall see later in this section, this change does not lead to an observable effect in the off-shell Higgs processes.

New states appearing in beyond SM scenarios can modify the running of the relevant gauge and Yukawa couplings. Generically, the beta function for a coupling $Q$ is given as
\begin{equation}
\beta_Q = \beta_Q^{\rm SM} + \sum_{\rm s: ~massive ~new ~states} \theta(\mu-M_s) (N_s \beta_{s,Q}^{\rm NP})\;,
\end{equation}
where $\beta_Q^{\rm SM}$ is the SM beta function, and $\beta_{s,Q}^{\rm NP}$ represents the contribution of a new heavy state $s$ of mass $M_s$, with $N_s$ number of degenerate degrees of freedom. The theta function encodes the fact that the effect of new heavy states is included in the RG running once the energy scale $\mu$ is above the threshold $M_s$. 

Large modifications to the running couplings compared to the SM case are however not expected in four-dimensional quantum field theories essentially due to the logarithmic nature of the running. Taking the example of the minimal supersymmetric extension of the Standard Model (MSSM), it is straightforward to include the leading MSSM contributions to the running of the gauge and Yukawa couplings. The one-loop beta functions of the gauge couplings are modified to 
\begin{align}
b_i^{\rm MSSM} = & (33/5,1,-3), 
\end{align}
while the $y_t$ running is now given by
\begin{align}
\frac{dy_t}{d t} = & \frac{y_t}{16\pi^2} \left(6y_t^2 - \frac{16}{3} g_3^2 - 3 g_2^2 - \frac{13}{15} g_1^2\right), ~~~~~~~~{\rm MSSM} 
\end{align}
We illustrate the running of $y_t$ in the MSSM by the green dashed curve in Fig.~\ref{fig:yt_SM}, for a common sparticle mass scale of $500$ GeV. It is observed that primarily due to the slower running of the strong coupling, the $y_t$ running is also slower in the MSSM compared to the SM scenario, and hence not observable in the off-shell Higgs processes at the LHC. 

A qualitatively different scenario however is obtained if there is a tower of new physics states modifying the RGEs, asymptotically leading to a power-law running of the Yukawa coupling~\cite{ED_RGE}. This four-dimensional description is equivalent to a theory with compactified flat extra space-like dimensions, with gauge and/or matter fields propagating in the higher-dimensional bulk.
To illustrate this, we consider two scenarios of compactified flat extra-dimensions~\cite{Appelquist:2000nn}: a 5D model with the extra-dimension compactified on an $S_1/Z_2$ orbifold, and a 6D model with the two extra dimensions compactified on a square $T^2/Z_2$ orbifold~\cite{Appelquist:2000nn,Appelquist:2001mj}. In both cases, we only consider the SM gauge fields to be propagating in the bulk, with the matter fields of the SM restricted to the brane~\cite{5D,6D}. The presence of color adjoint massive gauge fields, namely the KK-gluons, and their corresponding scalar fifth components would then modify the running of the strong coupling $\alpha_S$, which, in turn, would dominantly modify the running of the top quark Yukawa coupling $y_t$. The beta functions of the gauge couplings in such scenarios are given as:
\begin{align}
b_i^{\rm 5D} = & b_i^{\rm SM} + (S(t)-1) \times (1/10,-41/6,-21/2) \nonumber\\
b_i^{\rm 6D} = & b_i^{\rm SM} + (\pi S(t)^2-1) \times (1/10,-13/2,-10).
\end{align}
Here, $S(t)$ counts the number of degrees of freedom $S(t)=e^{t}R$, $R$ being the radius of the extra dimension.
The corresponding one-loop RGE equations for the Yukawa coupling in the extra-dimensional scenarios are as follows
\begin{align}
\frac{dy_t}{d t} = & \beta_{y_t}^{\rm SM}  +\frac{y_t}{16\pi^2} 2(S(t)-1)\left(\frac{3}{2} y_t^2 - 8 g_3^2 - \frac{9}{4} g_2^2 - \frac{17}{20} g_1^2\right), &{\rm 5D,} \nonumber \\
\frac{dy_t}{d t} = & \beta_{y_t}^{\rm SM}  +\frac{y_t}{16\pi^2} 4\pi(S(t)^2-1)\left(\frac{3}{2} y_t^2 - 8 g_3^2 - \frac{9}{4} g_2^2 - \frac{17}{20} g_1^2\right), &{\rm 6D}.
\end{align}
We see from Fig.~\ref{fig:yt_SM} that in the presence of such a tower of new states, the running of $y_t$ can be substantially altered for both the 5D  (blue dot-dashed line), and 6D (red dashed line) models.  

\begin{figure}[t!]
\centering
\includegraphics[scale=0.263]{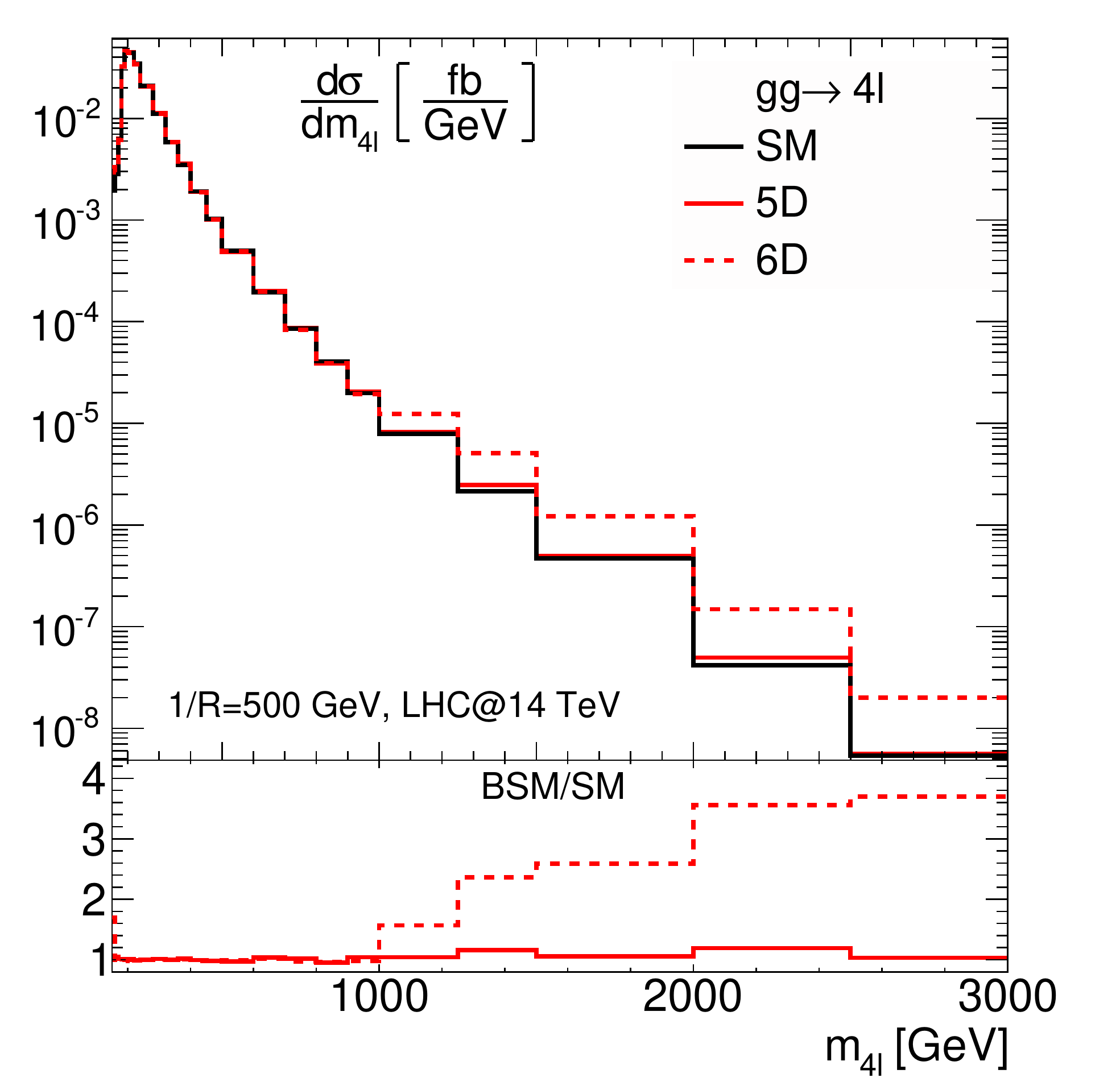} \hspace{-.6cm}
\includegraphics[scale=0.263]{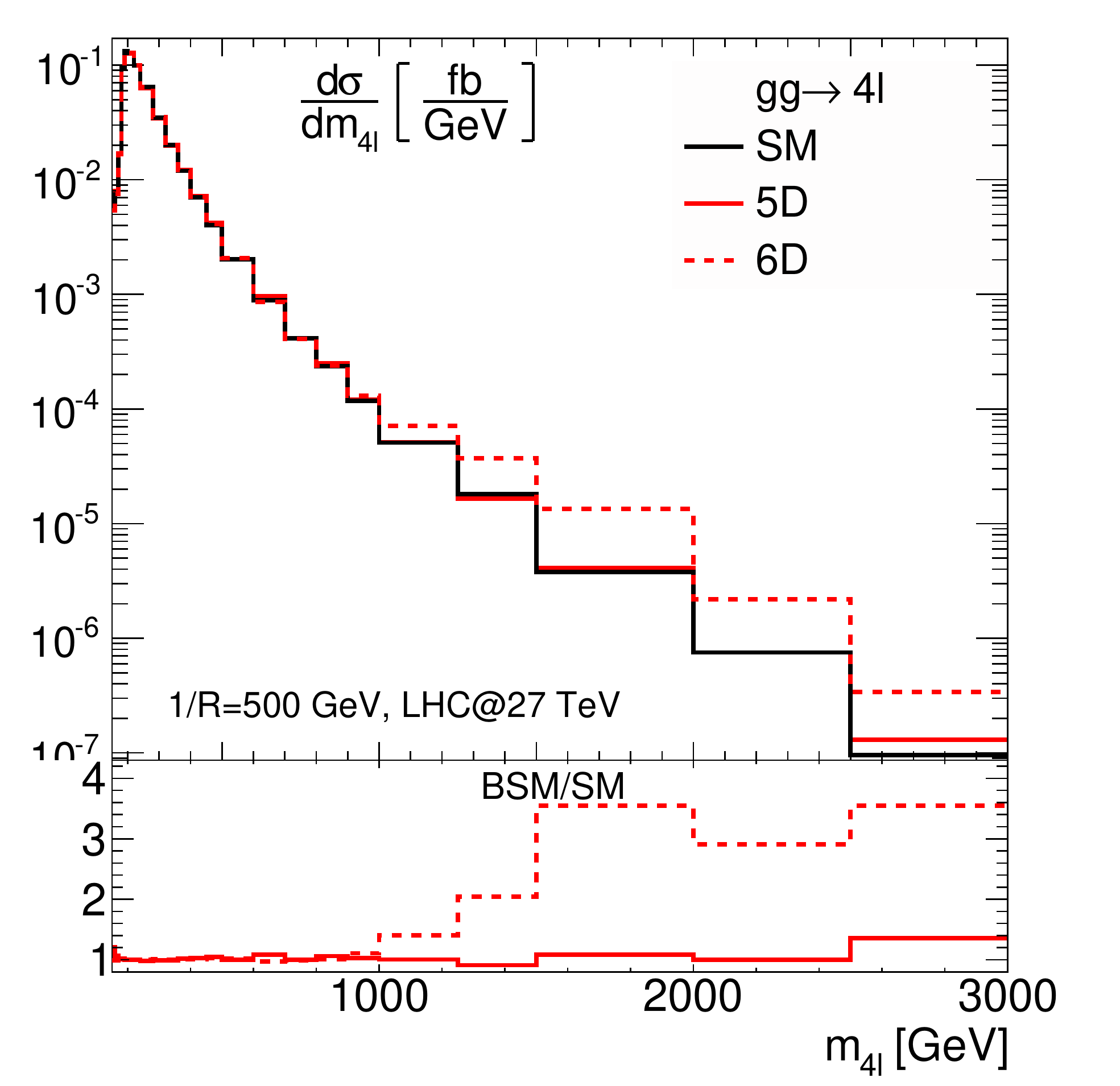} \hspace{-.6cm}
\includegraphics[scale=0.42]{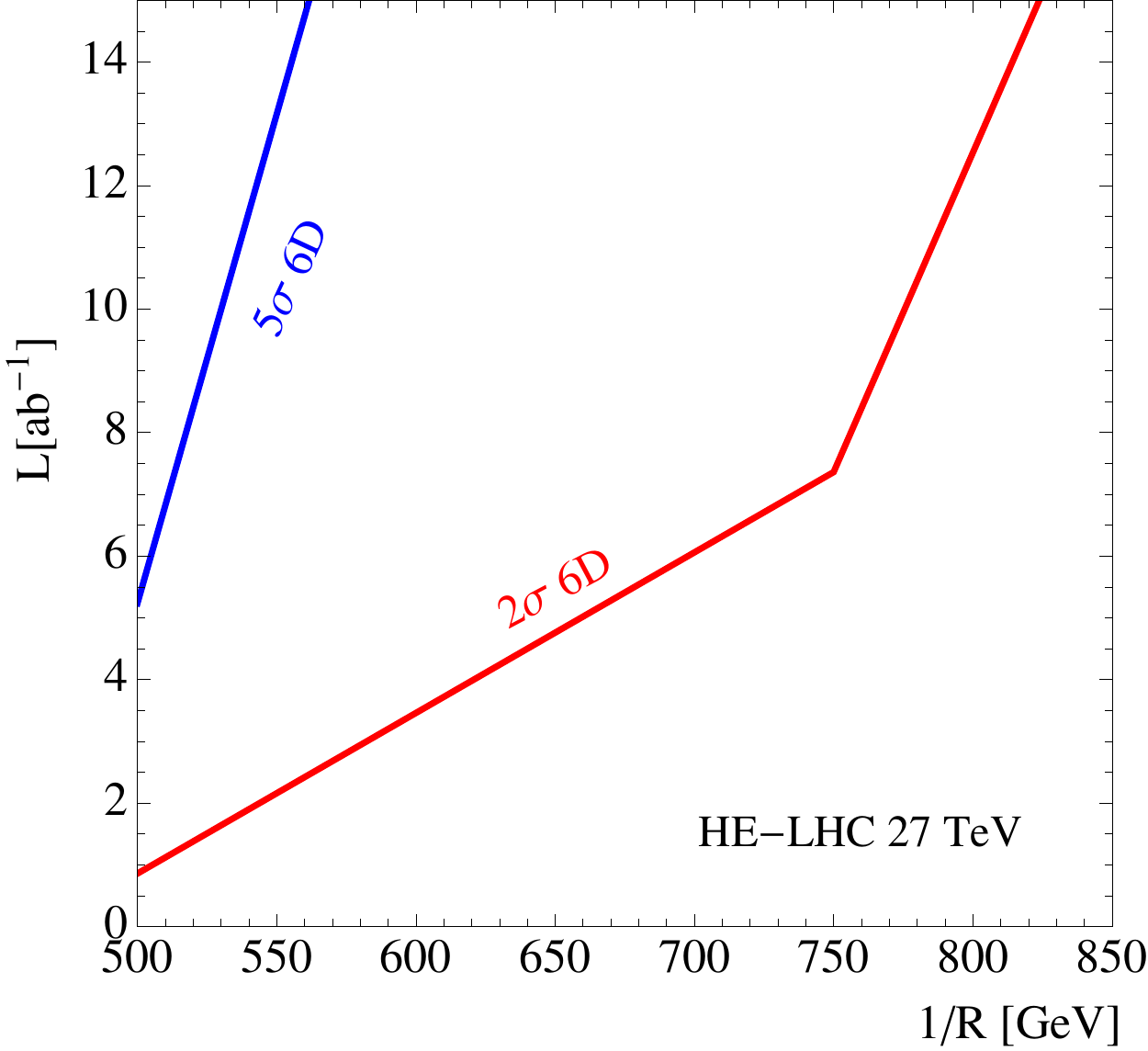}
\caption{
Four-lepton invariant mass distribution for the $gg\rightarrow 4\ell$ process at the LHC $14$~TeV (left) and 27 TeV (center) 
for the SM (black), and the 5D (red solid) and 6D (red dashed) extra-dimensional models, including the respective RGE evolution for $y_t$, at the $\sqrt{s}=14$~TeV LHC (left) and the 27~TeV HE-LHC (center), with the inverse radius $1/R=500$~GeV. 
We show the signal ratio between the extra-dimension model and the SM in the bottom panels.
Right: $2\sigma$ (red) and  $5\sigma$ (blue) bounds on the 6D model scale $1/R$ at the HE-LHC.}
\label{fig:m4l_UED}
\end{figure}

Following the analysis setup discussed in Sec.~\ref{sec:process}, we now describe the impact of the modified RG running in the $p p \rightarrow ZZ$ process. In Fig.~\ref{fig:m4l_UED} (left and center), we display the $m_{4\ell}$ distributions accounting for the top Yukawa $y_t$  RG evolution in the SM,
5D and 6D, assuming $1/R=500$~GeV for the latter two scenarios. 
We show the signal ratio between the extra-dimension model and the SM in the bottom panels, which could be upto a factor of four
in case of the 6D scenario. Although we do not observe relevant sensitivity to the 5D model due to the numerically less significant change with respect to the SM, in the 6D model we can probe $1/R\sim0.8$~TeV at the $2\sigma$ confidence level, with 15~ab$^{-1}$ of data at the HE-LHC. 

We note that although the impact of large deviations in the RGE running of $y_t$ can be clearly observed in the $p p \rightarrow ZZ$ process, this measurement alone is not sufficient to extract the value of running $y_t$ at higher scales. The latter interpretation would require the measurement of at least one other independent process at the LHC. This is because the running strong coupling $\alpha_S$ also enters all production processes at the LHC: through the hard scattering strong interaction, through parton shower evolution of the initial and final state quarks and gluons, and through the modification of the parton distribution functions (PDF). The PDFs are further modified by the addition of new splitting amplitudes of the gluon, thereby altering the DGLAP evolution equations. Therefore, for a complete experimental understanding of the RG evolution of different couplings in an extra-dimensional scenario, we first need to determine the modifications in $\alpha_S$ and the PDFs from multi-jet production at the LHC, in particular, from the ratio of two and three jet cross-sections. Subsequently, we can utilize the $p p \rightarrow h^* \rightarrow ZZ$ production to extract the information on running of $y_t$. 

The six extra-dimension scenario, which shows promising sensitivity in off-shell Higgs measurements above, can also be probed using the direct production of KK-gauge bosons at the LHC and its upgrades. Among these states, the KK-gluons have the highest production rate. Depending upon the details of the model realization, either both the even and the odd KK-states, or only the even KK-states, would have couplings to a pair of SM fermions. Therefore, the best probe at the LHC for the KK-gluons, for example, would be a dijet resonance search, the bounds from which depend on its coupling to the SM particles, the mass scale and the width of the lowest-lying KK states, and can become competitive or better than the off-shell Higgs probe discussed above~\cite{UED_LHC}. However, as emphasized earlier, the off-shell Higgs process constitutes a model-independent probe of such states, which can give a hint to the presence of a tower of states from a single measurement.

\section{Strongly Coupled Scenario: Form Factor}
\label{sec:composite}

Although the observed properties of the 125 GeV Higgs boson are consistent with the SM prediction of an elementary scalar Higgs doublet, given the present accuracy of the LHC measurements, it remains an open possibility that the Higgs boson is composite in nature, being a bound state of a confining strongly interacting theory with a characteristic compositeness scale of $\Lambda$. At the same time, the heaviest fermion in the SM, namely the top quark could be composite (or partially composite) as well. In this section we shall discuss some generic expectations for such a scenario.

Assuming parity conservation, and restricting ourselves to dimension-four couplings, generically the top-Higgs coupling will then involve a momentum-dependent form factor which is a function of all the independent Lorentz invariant combinations of the top ($p^\mu$) and the anti-top four-momenta ($\bar{p}^\mu$). Normalizing to the SM coupling, the off-shell top-Higgs effective vertex is then given as
\begin{equation}
V_{ttH}(p^\mu, \bar{p}^\mu) = \frac{\sqrt{2}m_t}{v} \Gamma \left(p^2/\Lambda^2, \bar{p}^2/\Lambda^2, q^2/\Lambda^2 \right), 
\label{eq:gen}
\end{equation}
where the Higgs boson four-momentum is given by $q^\mu = (p+\bar{p})^\mu$. 
In the limit $\Lambda \rightarrow \infty$, both the Higgs and the top are point-like particles, and therefore in this limit $\Gamma(0,0,0)=1$. 

Although the general form of such a three-point function is difficult to determine in a strongly interacting theory, one can gain an understanding of a composite scenario either in the large-${\rm N}$ limit (with ${\rm N}$ being the number of colors in a strongly coupled ${\rm SU} {\rm(N)}$ gauge theory)~\cite{Witten}, or within an weakly interacting warped five-dimensional model, which is dual to the four-dimensional strongly interacting theory in the large-N limit~\cite{Agashe}. While some aspects of such a scenario have been discussed in the literature~\cite{strong}, here we focus on a phenomenological ansatz for the form factor, which can be used to parametrize the expected deviations from the SM. 

In analogy with the nucleon electromagnetic form-factors \cite{FF,Godbole_Roy}, we adopt the following ansatz for the Higgs-top coupling form-factor:
\begin{equation}
\Gamma \left(q^2/\Lambda^2 \right) = \frac{1}{\left(1 + q^2/\Lambda^{2}\ \right)^{n}},
\label{eq:form}
\end{equation}
where $n=2$ corresponds to the dipole form-factor in the case of proton. As a large part of the total off-shell Higgs rate comes from the regime in which the top quarks in the triangle loop go on-shell, to simplify our analysis setup, we have set $p^2 =\bar{p}^2=m_t^2$ in the general form-factor in Eq.~\eqref{eq:gen}, thereby making it only a function of $q^2$. 

Since the on-shell couplings of the Higgs boson, and in particular the signal strength in the $ZZ^{(*)}$ final state is now well-measured to an accuracy of $\mathcal{O}(10\%)$, and since the measurement in this final state is driven by the gluon fusion production, the above form factors will be further constrained for $q^2=m_h^2$. In order to satisfy the on-shell Higgs constraints, we demand that  
\begin{equation}
|\Gamma \left(m_h^2/\Lambda^2 \right)^2-1| < 0.1
\end{equation}
at $95\%$ C.L. 
\begin{figure}[t]
\centering
    \includegraphics[width=0.44\textwidth]{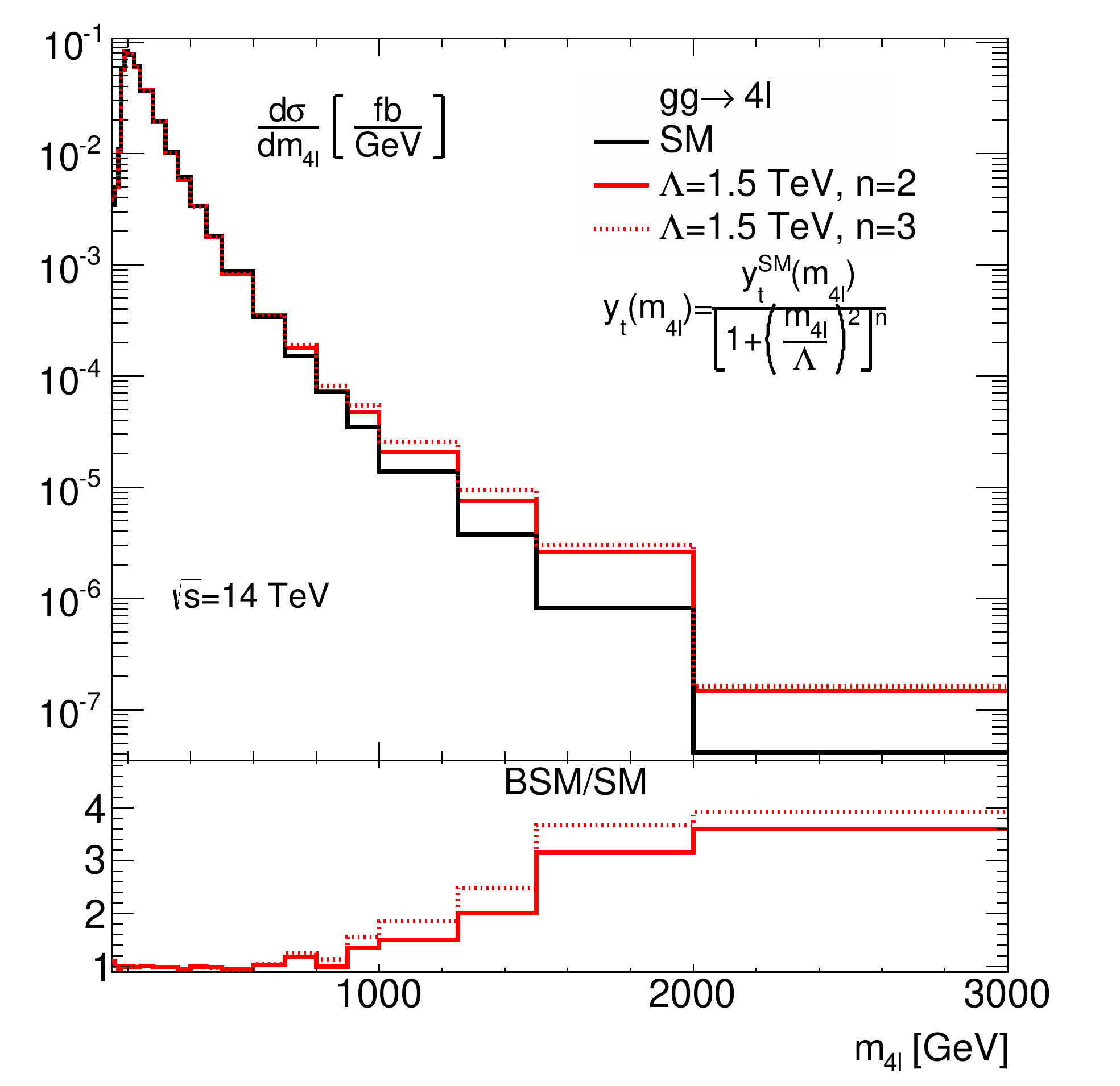}\hspace{0.1cm}
    \includegraphics[width=0.44\textwidth]{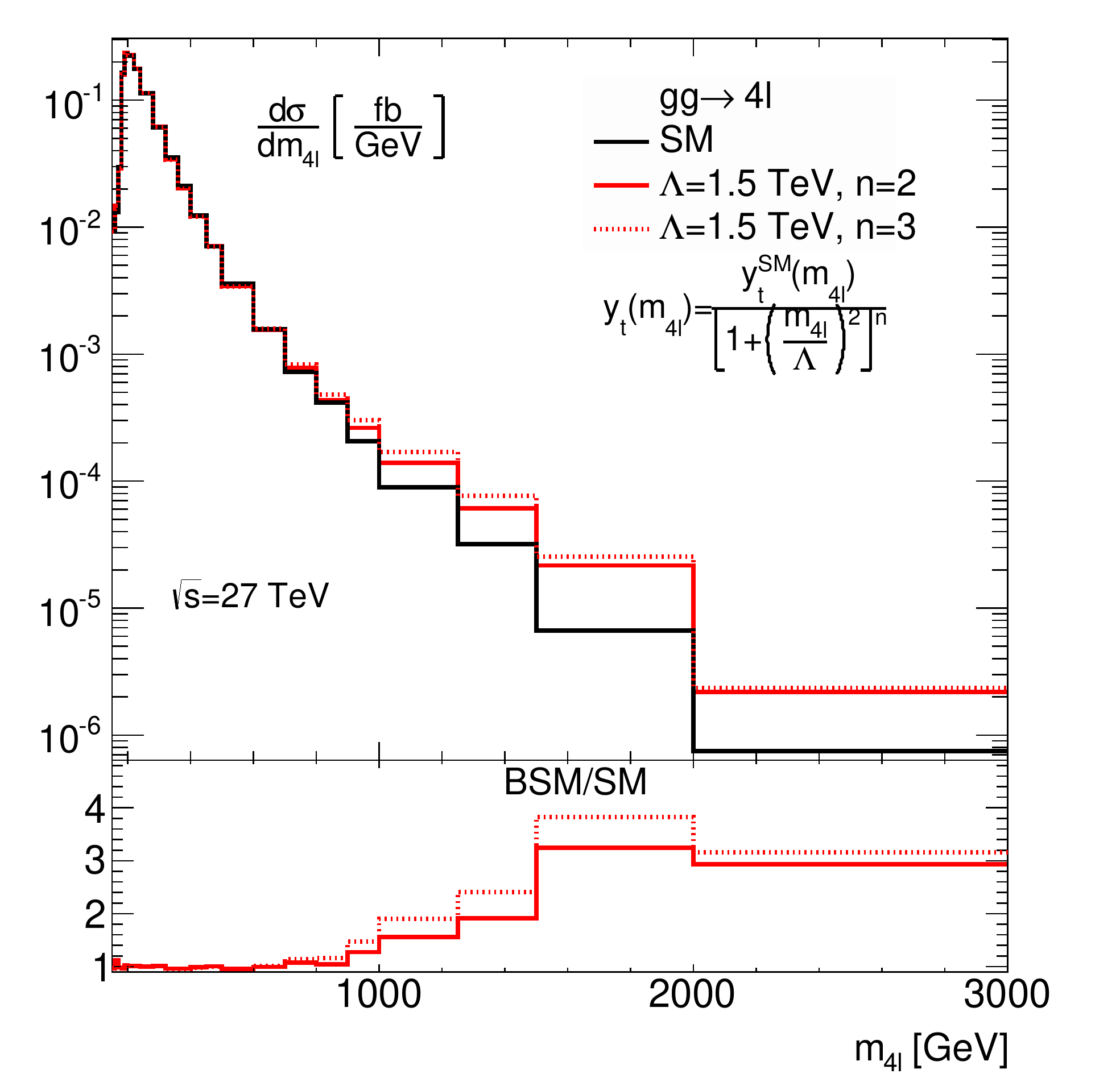}\\
    \includegraphics[width=0.4\textwidth]{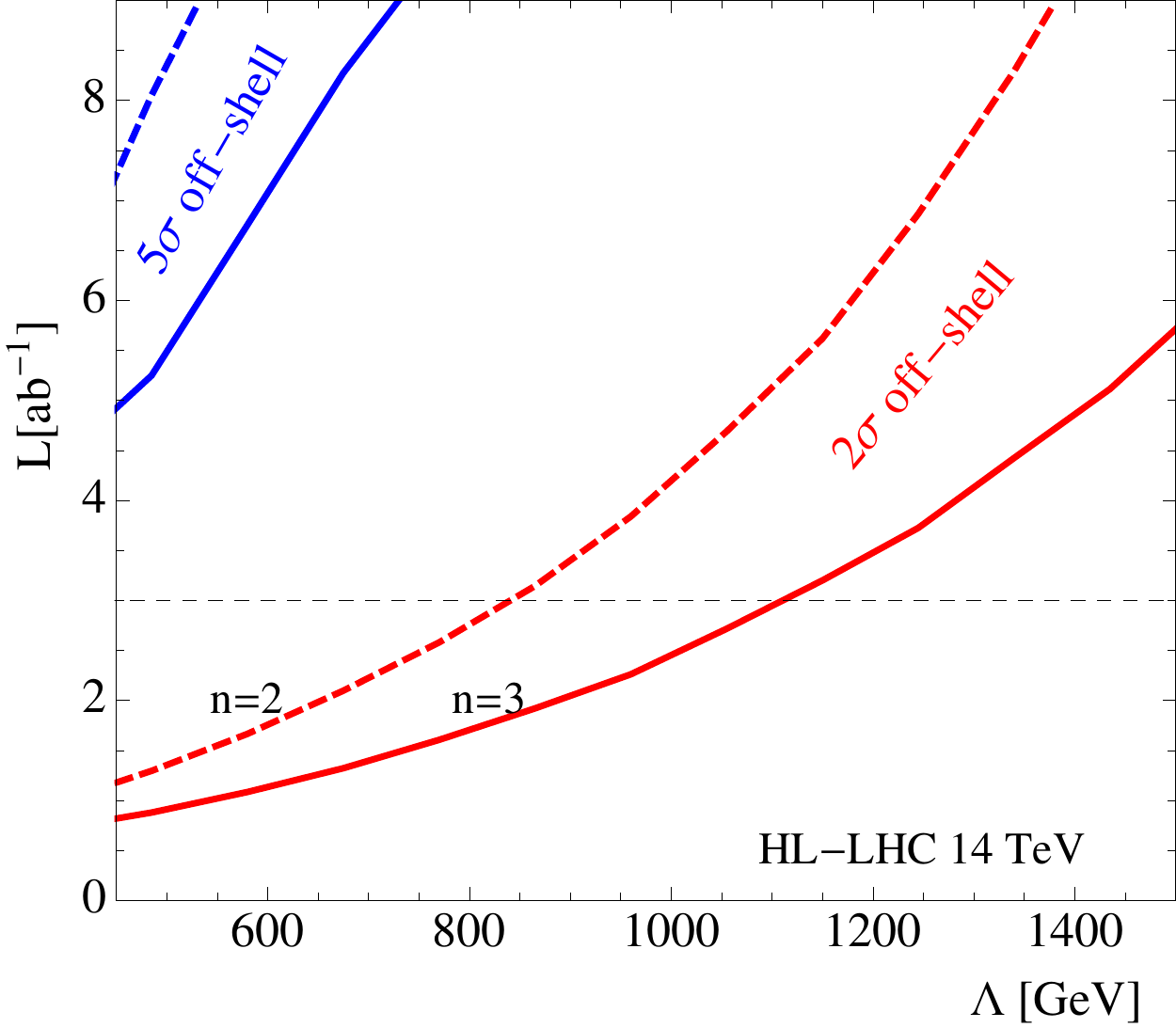}\hspace{0.3cm}
    \includegraphics[width=0.425\textwidth]{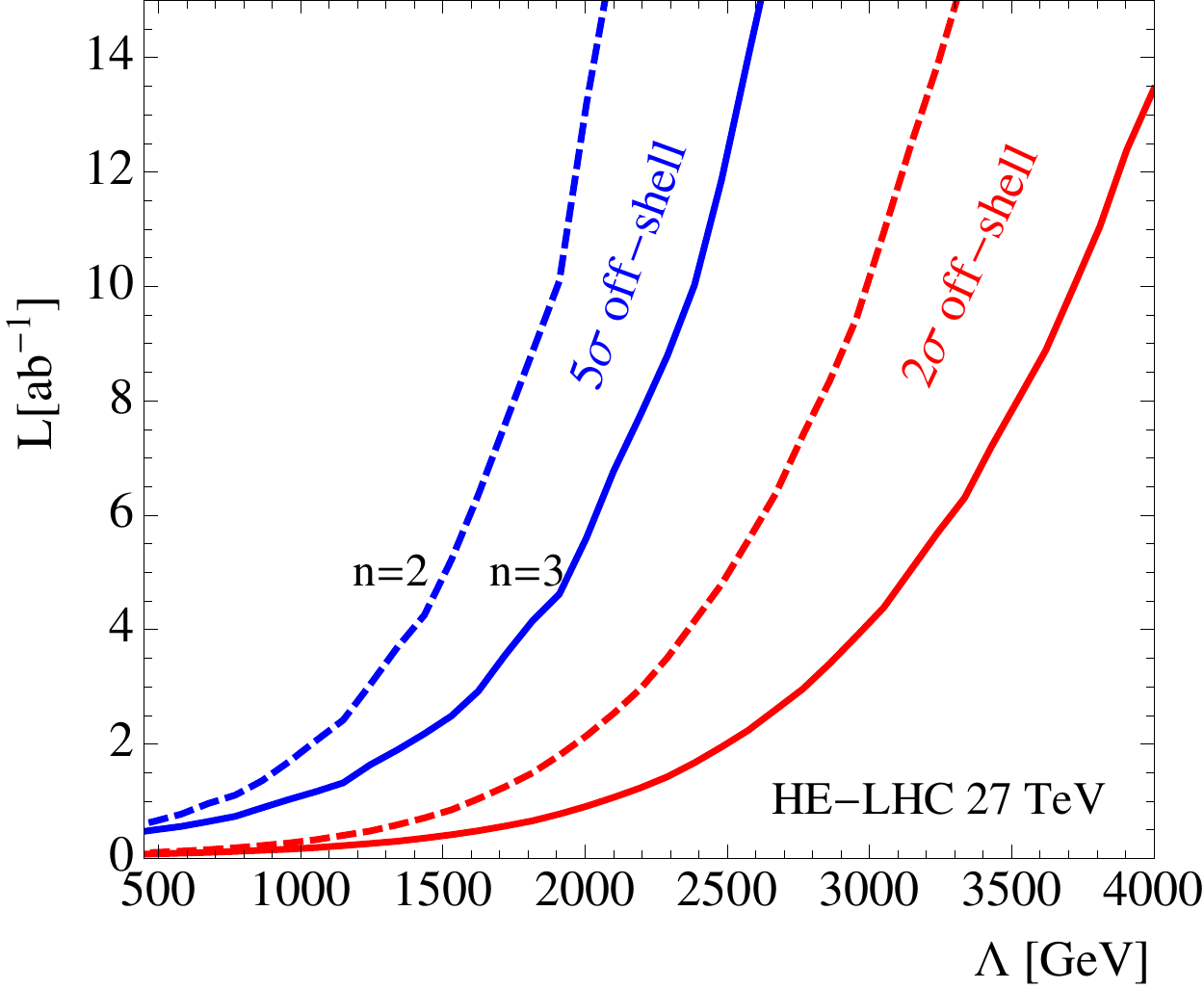}
 \caption{Top: Four-lepton invariant mass distribution for the $gg\rightarrow 4\ell$ process at the LHC $14$~TeV (upper left) and 27 TeV (upper right) in the SM (black) and in the presence of a form-factor in the Higgs-top coupling for  strong interaction scale $\Lambda=1.5$~TeV (red). We show the signal ratio between the form factor model and the SM in the bottom panels.
 Bottom: $5\sigma$ (blue) and $2\sigma$ (red) bounds on the scale $\Lambda$, for two different form factor scenarios with $n=2$ (dashed) and $n=3$ (solid), at   the $14$~TeV (left) and 27~TeV (right) LHC.
 }
 \label{fig:m4l_FF}
\end{figure}

There are different regimes of the energy scale $q^2$ for which a form-factor can be used to parametrize the underlying physics process. For $q^2<\Lambda^2$, the form factor can capture both semi-perturbative physics, {\it e.g.}, top-partner and top quark mixing in composite Higgs scenarios (where $\Lambda$ is the mass-scale of the top partners)~\cite{strong}, as well as the generic effect of a finite-sized composite Higgs boson (where $\Lambda$ is the strong interaction scale above which the constituents of the Higgs would enter the complete description of the physics process). However, in analogy with elastic nucleon scattering at energies larger than $\mathcal{O}(1)$~GeV, even for $q^2>\Lambda^2$, a part of the total $g g \rightarrow ZZ$ cross-section stems from scattering processes where the Higgs boson is still the relevant degree of freedom, and therefore the form-factor description with an interaction of the form Eq.~\eqref{eq:gen} holds. This would of course lead to a suppressed contribution from the Higgs diagram, as the total cross-section for $q^2>\Lambda^2$ is dominated by the ``deeply inelastic regime'' instead. Since such a scenario leads to a rather dramatic prediction observable in the near future LHC measurements, we adopt this for our illustration of the LHC observability. 

We show the impact of the form-factor in the top-Higgs Yukawa coupling in the differential distribution of $m_{4\ell}$ for the $g g \rightarrow ZZ$ process in Fig.~\ref{fig:m4l_FF} (upper panels), for the choice of the compositeness scale $\Lambda=1.5$~TeV. The results are shown for the 14 TeV HL-LHC (upper left) and the 27 TeV HE-LHC upgrade (upper right), whereby we compare the SM prediction (solid black) and the prediction for different values of the form-factor exponent, as defined in Eq.~\eqref{eq:form} (solid red for $n=2$ and dotted red for $n=3$). The choice $n=2$ is in analogy with the proton electromagnetic dipole form-factor \cite{FF},  and the choice $n=3$ is representative of higher multipoles. We show the signal ratio between the form factor model and the SM in the bottom panels, which can be upto a factor of four.
As we can see from Fig.~\ref{fig:m4l_FF} (lower panels), with the accumulation of $15 {~\rm ab}^{-1}$ of data, the HE-LHC can probe, at the $5\sigma$ C.L., values of the compositeness scale upto $\Lambda=2$~TeV for $n=2$, and $\Lambda=$2.5~TeV for $n=3$, exploring the off-shell Higgs measurement. Remarkably, these values of $\Lambda$ correspond to deviations in the top Yukawa below $\mathcal{O}(1 \%)$ for  on-shell Higgs production, which are very challenging to probe. Hence, the off-shell profile measurement is crucial to determine the presence of such a form-factor.
 
As mentioned above, we note that there are concrete realizations of the composite Higgs in which a semi-perturbative treatment of the low-energy effective theory can be realized, with the Higgs being a pseudo-Nambu-Goldstone boson of a spontaneously broken global symmetry~\cite{strong}. In addition, for a large class of scenarios, the top quark is partially composite in nature, in which elementary top quark fields mix linearly with composite fermionic operators. The latter also gives rise to a set of top partner fields. On integrating out the top-partner states, the top-Higgs Yukawa coupling receives a momentum-dependent form-factor correction, where the poles of the form-factors correspond to the fermionic resonance masses~\cite{strong, top-partner}. Thus the above general description of form-factors can also be applied to such composite Higgs scenarios, as long as the relevant momentum scales involved are smaller than the top-partner mass scale.

\section{Summary}
\label{sec:summary}
Following the Higgs boson discovery at the LHC, the major focus of the current experimental programme is to determine both the strength and the Lorentz structure of Higgs couplings to different SM particles, with the experimental search criteria chosen to probe  on-shell Higgs production and decays in most cases. To explore the answer to the ``naturalness problem'' at higher scales, a next step would be to study the energy scale dependence of the Higgs processes in general, and of Higgs couplings in particular, utilizing the off-shell production of the Higgs boson. In this connection, it is encouraging to note that off-shell Higgs production in weak-gauge boson pair final states has a significant event rate at the LHC and at higher energy hadron colliders, as shown in Sec.~\ref{sec:process}, thus offering a potential probe of Higgs couplings to top quarks and weak bosons at higher scales. 

In this paper, we studied several representative scenarios in which the Higgs production processes and Higgs couplings show significant dependence on the energy scale of the process, and explored what we can learn about such scenarios utilizing hadron colliders in the near future. We emphasized that a study of the off-shell Higgs process may shed light on new physics scenarios associated with naturalness of the Higgs boson mass. In the absence of signals for new physics in conventional direct searches at the LHC, our proposal constitutes a rather conservative approach with a broad applicability towards exploring the ultraviolet regime. 

Most scenarios beyond the SM, when invoked to address the question of stabilizing the Higgs boson mass against large radiative corrections from the next energy scale, not only predict deviations in Higgs total rates and differential distributions of interest, but also often give rise to new states around the electroweak scale. If the production rates and decay topologies of these new particles are favorable, they can be probed through direct production at colliders. However, the new states may not carry SM gauge charges (as in twin Higgs and some neutral naturalness scenarios), or may be difficult to observe in direct production due to low visible energy in the final states. Whichever may be the case, we found that such new particles should always show up as momentum dependent radiative corrections in off-shell Higgs production. This is best illustrated through a new gauge singlet scalar sector coupled to the Higgs, and we observed that the off-shell Higgs probe presented leads to improved reach in the singlet sector parameter space, compared to direct production of scalar singlet pair in weak-boson fusion, as demonstrated in Sec.~\ref{sec:portal}.  We found that  for values of the portal coupling implied by the naturalness relation (at the one-loop order), the $27$~TeV HE-LHC upgrade can probe singlet scalar masses of around $120$~GeV at the $2\sigma$ confidence. Our results are summarized in Table \ref{tab:sum}.

To push the branch-cut contribution to an extreme, we adopt the formulation of the ``quantum critical Higgs'' scenario (QCH), as discussed in Sec.~\ref{sec:conformal}, where the new physics threshold comes in as a continuum spectrum. With the formalism as in \cite{Bellazzini:2015cgj}, 
we found significant sensitivity to the new scale parameter, reaching a possible $5\sigma$ observation for $\mu\sim 0.9$ TeV at the HE-LHC.

Although the heart of the ``naturalness problem'' with the quadratic sensitivity to new physics must be deeply rooted in the ultraviolet, the scale evolution of the Higgs couplings is logarithmic in nature if the physics is indeed governed by a renormalizable four-dimensional quantum field theory. We studied in Sec.~\ref{sec:RGE} scenarios in which the renormalization group evolution of Higgs couplings can be modified compared to their SM expectation. In particular, we focussed on the RG running of the top quark Yukawa coupling $-$ in the SM, MSSM, and in extra-dimensional scenarios. In the first two cases, the logarithmic running between the TeV scale and the electroweak scale is well predicted, but not large enough to have an observable impact at currently planned experiments, while in the presence of large flat extra-dimensions the running of the top Yukawa can be significantly modified compared to its SM expectation. The key point in this regard is that the contributions from an equally spaced tower of KK resonances can lead to an asymptotically power-law running of the gauge and Yukawa couplings. Taking an example scenario with only the SM gauge fields propagating in the bulk, while the matter fields are restricted to the brane, we observed that with one extra space-like dimension, the modification with respect to the SM running is numerically not significant enough to be observable at the LHC or the HE-LHC upgrade. On the other hand, with two extra-dimensions, the sensitivity reaches $2\sigma$ confidence level for $1/R \sim 0.8$ TeV with $15 {~\rm ab}^{-1}$ data at the HE-LHC. We note that although the particular flat extra-dimensional scenario adopted does not have a direct implication for naturalness, the framework is illustrative of how different the largest Higgs coupling in the SM can become in the ultraviolet.

\begin{table}[t]
\centering
\begin{tabular}{|c|c|c|c|c|}
\hline 
 & Singlet & QCH ($\Delta= 1.1$) & Two Extra-Dim. & Form Factor ($n=2$) \\ 
\hline 
HL-LHC, 2$\sigma$ & $m_S\sim 70$ GeV &  $\mu \sim 0.5$~TeV &  $-$  & $\Lambda \sim 0.8$ TeV  \\ 
\hline 
HE-LHC, 2$\sigma$ & $m_S\sim 120$ GeV &    $\mu \sim 1.6$~TeV   & $R^{-1}\sim 0.8$~TeV & $\Lambda \sim 3.3$ TeV \\ 
5$\sigma$ reach & $m_S\sim 100$ GeV  & $\mu \sim  0.9$~TeV  & $R^{-1}\sim 0.6$~TeV & $\Lambda \sim 2.1$ TeV \\ 
\hline 
\end{tabular}
\caption{
\label{tab:sum}
{The reach in the mass scale and model parameters for various theoretical scenarios at the HL-LHC (14 TeV, 3 ab$^{-1}$) and the HE-LHC (27 TeV, 15 ab$^{-1}$).}}
\end{table} 

As a final example, we studied certain generic implications of the Higgs boson being a composite state of a new strongly interacting sector in Sec.~\ref{sec:composite}. Once again focussing on the top-Higgs sector, we introduced a form-factor in the effective Yukawa vertex, which parametrizes the composite nature of the Higgs particle, leading to deviations from the SM predictions at higher momentum transfers. To illustrate the phenomenological impact of such a form factor in off-shell Higgs production, we adopted an ansatz in analogy with the proton electromagnetic form factor. At energies above the compositeness scale, the form-factor description is applicable to processes in which the Higgs boson is still the relevant degree of freedom, with a suppressed contribution to the total rate.  In such a regime with a high momentum transfer, the probability for the Higgs to be intact is very low and the dominant events would be deeply inelastic, with direct production of the constituents, which we chose not to quantify. We found that the shape of the differential distribution of the $Z$ boson pair invariant mass in the off-shell region to be highly sensitive to the presence of a form-factor. Consequently, even for compositeness scales $\Lambda$ that lead to less than $\mathcal{O}(1\%)$ deviations in on-shell rates, the off-shell measurements can be utilized as a promising probe. For values of the form factor exponent $n=2$ ($n=3$) we observe that scales $\Lambda$ around $2.1$~TeV ($2.5$~TeV) can be thus probed at the $5\sigma$ level using $15 {~\rm fb}^{-1}$ data at the HE-LHC . 

In closing, we argued that it is a natural and necessary next step to explore the Higgs physics at higher energy scales. To this end, our present work has demonstrated some meaningful example scenarios, which can provide a general roadmap in seeking for a possible solution to the ``naturalness problem''. Much work still needs to be done, including exploiting other channels for the Higgs production and decay, and other theoretical ideas on naturalness which are manifested in off-shell Higgs production at higher energies.

\section*{Acknowledgment}
We thank Ali Shayegan Shirazi and John Terning for helpful  discussions on the Quantum Critical Higgs framework, and for providing us with their numerical implementation of the relevant formalism for cross-checks. TH would like to thank the CERN Theoretical Physics Department for hospitality during the final stage of this project, and thank Gian Giudice and Matthew McCullough for discussions. This work was supported by the U.S.~Department of Energy under grant No.~DE-FG02- 95ER40896 and by the PITT PACC. DG is also supported by the U.S.~National Science Foundation under the grant PHY-1519175.


\begin{thebibliography}{99}
\bibitem{Goncalves:2017iub} 
  D.~Goncalves, T.~Han and S.~Mukhopadhyay,
  Phys.\ Rev.\ Lett.\  {\bf 120}, 111801 (2018)
 [arXiv:1710.02149 [hep-ph]].
  
\bibitem{Observations}
  G.~Aad {\it et al.} [ATLAS Collaboration],
  Phys.\ Lett.\ B {\bf 716}, 1 (2012) 
  [arXiv:1207.7214 [hep-ex]].
  S.~Chatrchyan {\it et al.} [CMS Collaboration],
  Phys.\ Lett.\ B {\bf 716}, 30 (2012).
\bibitem{Naturalness}
  E.~Gildener and S.~Weinberg,
  Phys.\ Rev.\ D {\bf 13}, 3333 (1976);
  G.~'t Hooft,
  NATO Sci.\ Ser.\ B {\bf 59}, 135 (1980).

\bibitem{susy}
For a recent review, see, for example,
  N.~Craig,
  arXiv:1309.0528 [hep-ph].
  
\bibitem{strong}
For a review, see, for example,
  G.~Panico and A.~Wulzer,
  Lect.\ Notes Phys.\  {\bf 913}, pp.1 (2016)
  [arXiv:1506.01961 [hep-ph]].
  
\bibitem{Csaki:2015hcd} 
  C.~Csaki, C.~Grojean and J.~Terning,
  Rev.\ Mod.\ Phys.\  {\bf 88}, no. 4, 045001 (2016)
  [arXiv:1512.00468 [hep-ph]].
  
\bibitem{Bellazzini:2015cgj} 
  B.~Bellazzini, C.~Cs\'{a}ki, J.~Hubisz, S.~J.~Lee, J.~Serra and J.~Terning,
  Phys.\ Rev.\ X {\bf 6}, no. 4, 041050 (2016).


  
\bibitem{ED_RGE}
  K.~R.~Dienes, E.~Dudas and T.~Gherghetta,
  Phys.\ Lett.\ B {\bf 436}, 55 (1998)
  [hep-ph/9803466]; 
  K.~R.~Dienes, E.~Dudas and T.~Gherghetta,
  Nucl.\ Phys.\ B {\bf 537}, 47 (1999)
  [hep-ph/9806292].
  
\bibitem{Appelquist:2000nn} 
  T.~Appelquist, H.~C.~Cheng and B.~A.~Dobrescu,
  Phys.\ Rev.\ D {\bf 64}, 035002 (2001)
  doi:10.1103/PhysRevD.64.035002
  [hep-ph/0012100].

\bibitem{Appelquist:2001mj} 
  T.~Appelquist, B.~A.~Dobrescu, E.~Ponton and H.~U.~Yee,
  Phys.\ Rev.\ Lett.\  {\bf 87}, 181802 (2001)
  doi:10.1103/PhysRevLett.87.181802
  [hep-ph/0107056].

 \bibitem{5D} 
  G.~Bhattacharyya, A.~Datta, S.~K.~Majee and A.~Raychaudhuri,
  Nucl.\ Phys.\ B {\bf 760}, 117 (2007);
  A.~S.~Cornell, A.~Deandrea, L.~X.~Liu and A.~Tarhini,
  Mod.\ Phys.\ Lett.\ A {\bf 28}, no. 11, 1330007 (2013);
  M.~Blennow, H.~Melbeus, T.~Ohlsson and H.~Zhang,
  JHEP {\bf 1104}, 052 (2011)
  [arXiv:1101.2585 [hep-ph]].

\bibitem{6D}
  T.~Kakuda, K.~Nishiwaki, K.~y.~Oda and R.~Watanabe,
  Phys.\ Rev.\ D {\bf 88}, 035007 (2013);
  A.~Abdalgabar, A.~S.~Cornell, A.~Deandrea and A.~Tarhini,
  Phys.\ Rev.\ D {\bf 88}, 056006 (2013)
  [arXiv:1306.4852 [hep-ph]].

 \bibitem{ATLAStth}
  M.~Aaboud {\it et al.} [ATLAS Collaboration],
  arXiv:1712.08891 [hep-ex].
  
  \bibitem{CMStth}
  CMS Collaboration [CMS Collaboration],
  CMS-PAS-HIG-17-004.

\bibitem{ATLASsigma}
  M.~Aaboud {\it et al.} [ATLAS Collaboration],
  arXiv:1712.02304 [hep-ex];
  The ATLAS collaboration [ATLAS Collaboration],
  ATLAS-CONF-2017-047.

\bibitem{CMSzzh}
  CMS Collaboration [CMS Collaboration],
  CMS-PAS-HIG-16-041.

\bibitem{ATLASwwh}
  G.~Aad {\it et al.} [ATLAS Collaboration],
  JHEP {\bf 1608}, 104 (2016);
  The ATLAS collaboration [ATLAS Collaboration],
  ATLAS-CONF-2016-112.

\bibitem{CMSwwh}
  CMS Collaboration [CMS Collaboration],
  CMS-PAS-HIG-16-021.

\bibitem{ATLAStau}
  G.~Aad {\it et al.} [ATLAS Collaboration],
  JHEP {\bf 1504}, 117 (2015)
 [arXiv:1501.04943 [hep-ex]].

\bibitem{CMStau}
  CMS Collaboration [CMS Collaboration],
  CMS-PAS-HIG-16-043.

\bibitem{ATLASbb}
  M.~Aaboud {\it et al.} [ATLAS Collaboration],
  JHEP {\bf 1712}, 024 (2017)
  [arXiv:1708.03299 [hep-ex]].
  
\bibitem{CMSbb}
  CMS Collaboration [CMS Collaboration],
  CMS-PAS-HIG-16-044.
  
  \bibitem{offshell} 
  N.~Kauer and G.~Passarino,
  JHEP {\bf 1208}, 116 (2012)
  [arxiv{1206.4803} [hep-ph]];
  F.~Caola and K.~Melnikov,
  Phys.\ Rev.\ D {\bf 88}, 054024 (2013)
  [arxiv{1307.4935} [hep-ph]];
  J.~M.~Campbell, R.~K.~Ellis and C.~Williams,
  JHEP {\bf 1404}, 060 (2014)
  [arxiv{1311.3589} [hep-ph]].

  \bibitem{offshell_ex}
  CMS Collaboration, CMS-PAS-HIG-13-002 (2013);
  CMS Collaboration,
  Phys.\ Lett.\ B {\bf 736}, 64 (2014),
  [arxiv{1405.3455} [hep-ex]];
  ATLAS collaboration,
  ATLAS-CONF-2014-042.

\bibitem{offshell_model} 
  C.~Englert and M.~Spannowsky,
  Phys.\ Rev.\ D {\bf 90}, 053003 (2014)
  [arxiv{1405.0285} [hep-ph]]; 
    G.~Cacciapaglia, A.~Deandrea, G.~Drieu La Rochelle and J.~B.~Flament,
  Phys.\ Rev.\ Lett.\  {\bf 113}, no. 20, 201802 (2014)
  [arxiv{1406.1757} [hep-ph]];
  A.~Azatov, C.~Grojean, A.~Paul and E.~Salvioni,
  Zh.\ Eksp.\ Teor.\ Fiz.\  {\bf 147}, 410 (2015)
  [J.\ Exp.\ Theor.\ Phys.\  {\bf 120}, 354 (2015)];
  [arxiv{1406.6338} [hep-ph]];
  C.~Englert, Y.~Soreq and M.~Spannowsky,
  JHEP {\bf 1505}, 145 (2015)
  [arXiv:1410.5440 [hep-ph]];
  M.~Buschmann, D.~Goncalves, S.~Kuttimalai, M.~Sch\"onherr, F.~Krauss and T.~Plehn,
  JHEP {\bf 1502}, 038 (2015)
  [arxiv{1410.5806} [hep-ph]].

\bibitem{legacy} 
  T.~Corbett, O.~J.~P.~Eboli, D.~Goncalves, J.~Gonzalez-Fraile, T.~Plehn and M.~Rauch,
  JHEP {\bf 1508}, 156 (2015)
  [arXiv:1505.05516 [hep-ph]].
  
  \bibitem{width} 
  ATLAS Collaboration [ATLAS Collaboration],
  ATL-PHYS-PUB-2015-024. 

  \bibitem{mcfm} 
  J.~M.~Campbell, R.~K.~Ellis and C.~Williams, {\tt http://mcfm.fnal.gov}.
  
   \bibitem{cteq} 
  J.~Pumplin, D.~R.~Stump, J.~Huston, H.~L.~Lai, P.~M.~Nadolsky and W.~K.~Tung,
  JHEP {\bf 0207}, 012 (2002)
  [hep-ph/0201195].
    
 \bibitem{inv_portal} 
   V.~Silveira and A.~Zee,
  Phys.\ Lett.\ B {\bf 161}, 136 (1985)
  J.~McDonald,
  Phys.\ Rev.\ D\ {\bf 50}, 3637  (1994)
  [arxiv{hep-ph/0702143}];
  C.~P.~Burgess, M.~Pospelov and T.~ter Veldhuis,
  Nucl.\ Phys.\ B\ {\bf 619}, 709  (2001)
  [arxiv{hep-ph/0011335}];
  R.~Schabinger and J.~D.~Wells,
  Phys.\ Rev.\ D {\bf 72}, 093007 (2005)
  [arxiv{hep-ph/0509209}];
  B.~Patt and F.~Wilczek,
  arxiv{hep-ph/0605188};
  R.~Barbieri, T.~Gregoire and L.~J.~Hall,
  arxiv{hep-ph/0509242};
  M.~H.~G.~Tytgat,
  PoSIDM\ {\bf 2010}, 126  (2011)
  [arxiv{1012.0576} [hep-ph]];
  C.~Englert, T.~Plehn, D.~Zerwas and P.~M.~Zerwas,
  Phys.\ Lett.\ B {\bf 703}, 298 (2011)
  [arxiv{1106.3097} [hep-ph]];
  M.~Pospelov and A.~Ritz,
  Phys.\ Rev.\ D {\bf 84}, 113001 (2011)
  [arxiv{1109.4872} [hep-ph]];
  X.~-G.~He and J.~Tandean,
  Phys.\ Rev.\ D\ {\bf 84}, 075018  (2011)
  [arxiv{1109.1277} [hep-ph]];
  P.~J.~Fox, R.~Harnik, J.~Kopp and Y.~Tsai,
  Phys.\ Rev.\ D {\bf 85}, 056011 (2012)
  [arxiv{1109.4398} [hep-ph]];
  I.~Low, P.~Schwaller, G.~Shaughnessy and C.~E.~M.~Wagner,
  Phys.\ Rev.\ D {\bf 85}, 015009 (2012)
  [arxiv{1110.4405} [hep-ph]];
  C.~Englert, T.~Plehn, M.~Rauch, D.~Zerwas and P.~M.~Zerwas,
  Phys.\ Lett.\ B {\bf 707}, 512 (2012)
  [arxiv{1112.3007} [hep-ph]];
  A.~Djouadi, O.~Lebedev, Y.~Mambrini and J.~Quevillon,
  Phys.\ Lett.\ B {\bf 709}, 65 (2012)
  [arxiv{1112.3299} [hep-ph]];
  B.~Batell, S.~Gori and L.~T.~Wang,
  JHEP {\bf 1206}, 172 (2012)
  [arxiv{1112.5180} [hep-ph]];
  S.~Baek, P.~Ko and W.~I.~Park,
  Phys.\ Rev.\ D {\bf 90}, no. 5, 055014 (2014)
  [arxiv{1405.3530} [hep-ph]].
   C.~W.~Chiang, M.~J.~Ramsey-Musolf and E.~Senaha,
  arXiv:1707.09960 [hep-ph].

\bibitem{cline}
  J.~M.~Cline, K.~Kainulainen, P.~Scott and C.~Weniger,
  Phys.\ Rev.\ D {\bf 88}, 055025 (2013)
  Erratum: [Phys.\ Rev.\ D {\bf 92}, no. 3, 039906 (2015)]
  [arXiv:1306.4710 [hep-ph]].

  \bibitem{curtin}
  D.~Curtin, P.~Meade and C.~T.~Yu,
  JHEP {\bf 1411}, 127 (2014)
  [arXiv:1409.0005 [hep-ph]].
 
\bibitem{Twin} 
  Z.~Chacko, H.~S.~Goh and R.~Harnik,
  Phys.\ Rev.\ Lett.\  {\bf 96}, 231802 (2006).

\bibitem{Folded_SUSY}
  G.~Burdman, Z.~Chacko, H.~S.~Goh and R.~Harnik,
  JHEP {\bf 0702}, 009 (2007)
  [hep-ph/0609152].

\bibitem{Orbifold}
  N.~Craig, S.~Knapen and P.~Longhi,
  Phys.\ Rev.\ Lett.\  {\bf 114}, no. 6, 061803 (2015);
  N.~Craig, S.~Knapen and P.~Longhi,
  JHEP {\bf 1503}, 106 (2015)
  [arXiv:1411.7393 [hep-ph]].
  
\bibitem{signals_twin} 
  D.~Curtin and C.~B.~Verhaaren,
  JHEP {\bf 1512}, 072 (2015)
  doi:10.1007/JHEP12(2015)072
  [arXiv:1506.06141 [hep-ph]].
  
%
  \bibitem{complex_mass} 
  A.~Denner and S.~Dittmaier,
  Nucl.\ Phys.\ Proc.\ Suppl.\  {\bf 160}, 22 (2006)
  [hep-ph/0605312].
  
  \bibitem{Han}
  V.~Barger, T.~Han, P.~Langacker, B.~McElrath and P.~Zerwas,
  Phys.\ Rev.\ D {\bf 67}, 115001 (2003)
  [hep-ph/0301097].
  
  \bibitem{Englert:2013tya} 
  C.~Englert and M.~McCullough,
  JHEP {\bf 1307}, 168 (2013)
  [arXiv:1303.1526 [hep-ph]].
  N.~Craig, C.~Englert and M.~McCullough,
  Phys.\ Rev.\ Lett.\  {\bf 111}, no. 12, 121803 (2013)
  [arXiv:1305.5251 [hep-ph]].

\bibitem{read} 
  A.~L.~Read,
  J.\ Phys.\ G {\bf 28}, 2693 (2002).

  \bibitem{Barbieri} 
  R.~Barbieri and A.~Strumia,
  Phys.\ Lett.\ B {\bf 462}, 144 (1999);
  R.~Barbieri and A.~Strumia,
  hep-ph/0007265.

\bibitem{WBF}
  O.~J.~P.~Eboli and D.~Zeppenfeld,
  Phys.\ Lett.\ B {\bf 495}, 147 (2000)
  [hep-ph/0009158].
  N.~Craig, H.~K.~Lou, M.~McCullough and A.~Thalapillil,
  JHEP {\bf 1602}, 127 (2016)
  [arXiv:1412.0258 [hep-ph]].
 
\bibitem{CMS:2017cwx} 
  CMS Collaboration [CMS Collaboration],
  CMS-PAS-FTR-16-002.
    
  \bibitem{Stancato:2008mp} 
  D.~Stancato and J.~Terning,
  JHEP {\bf 0911}, 101 (2009)
  [arXiv:0807.3961 [hep-ph]].
    
 \bibitem{unhiggs} 
  C.~Englert, M.~Spannowsky, D.~Stancato and J.~Terning,
  Phys.\ Rev.\ D {\bf 85}, 095003 (2012)
  [arXiv:1203.0312 [hep-ph]].
  C.~Englert, D.~Goncalves, M.~Spannowsky and J.~Terning,
  Phys.\ Rev.\ D {\bf 86}, 035010 (2012)
  [arXiv:1205.0836 [hep-ph]].
  
   \bibitem{PDG}
  C.~Patrignani {\it et al.} [Particle Data Group],
  Chin.\ Phys.\ C {\bf 40}, no. 10, 100001 (2016);
For a recent measurement at the LHC, see, for example,
  M.~Aaboud {\it et al.} [ATLAS Collaboration],
  Eur.\ Phys.\ J.\ C {\bf 77}, no. 12, 872 (2017) 
  [arXiv:1707.02562 [hep-ex]].

\bibitem{Schwartz}
  D.~E.~Kaplan and M.~D.~Schwartz,
  Phys.\ Rev.\ Lett.\  {\bf 101}, 022002 (2008)
  [arXiv:0804.2477 [hep-ph]].

\bibitem{Michael}
  D.~Becciolini, M.~Gillioz, M.~Nardecchia, F.~Sannino and M.~Spannowsky,
  Phys.\ Rev.\ D {\bf 91}, no. 1, 015010 (2015)
  Addendum: [Phys.\ Rev.\ D {\bf 92}, no. 7, 079905 (2015)] 
  [arXiv:1403.7411 [hep-ph]].

\bibitem{Ruderman}  
  D.~S.~M.~Alves, J.~Galloway, J.~T.~Ruderman and J.~R.~Walsh,
  JHEP {\bf 1502}, 007 (2015) 
  [arXiv:1410.6810 [hep-ph]].
  
  
 \bibitem{UED_LHC}
  H.~C.~Cheng, K.~T.~Matchev and M.~Schmaltz,
  Phys.\ Rev.\ D {\bf 66}, 056006 (2002)
  [hep-ph/0205314];
  A.~Datta, K.~Kong and K.~T.~Matchev,
  Phys.\ Rev.\ D {\bf 72}, 096006 (2005)
  Erratum: [Phys.\ Rev.\ D {\bf 72}, 119901 (2005)]
  [hep-ph/0509246];
  G.~Burdman, O.~J.~P.~Eboli and D.~Spehler,
  Phys.\ Rev.\ D {\bf 94}, no. 9, 095004 (2016)
  [arXiv:1607.02260 [hep-ph]].

  
 \bibitem{Witten}
  E.~Witten,
  Nucl.\ Phys.\ B {\bf 160}, 57 (1979).

 
 \bibitem{Agashe} 
  K.~Agashe, R.~Contino and A.~Pomarol,
  Nucl.\ Phys.\ B {\bf 719}, 165 (2005)
  [hep-ph/0412089].


\bibitem{FF}
For a recent review, see, 
  V.~Punjabi, C.~F.~Perdrisat, M.~K.~Jones, E.~J.~Brash and C.~E.~Carlson,
  Eur.\ Phys.\ J.\ A {\bf 51}, 79 (2015)
  [arXiv:1503.01452 [nucl-ex]].
  
\bibitem{Godbole_Roy}
  R.~M.~Godbole and P.~Roy,
  Phys.\ Rev.\ Lett.\  {\bf 50}, 717 (1983).
  
\bibitem{top-partner}  
See, for example, 
  A.~Pomarol and F.~Riva,
  JHEP {\bf 1208}, 135 (2012)
  [arXiv:1205.6434 [hep-ph]];
  D.~Marzocca, M.~Serone and J.~Shu,
  JHEP {\bf 1208}, 013 (2012)
  [arXiv:1205.0770 [hep-ph]];
  G.~Panico and A.~Wulzer,
  JHEP {\bf 1109}, 135 (2011)
  [arXiv:1106.2719 [hep-ph]];
  O.~Matsedonskyi, G.~Panico and A.~Wulzer,
  JHEP {\bf 1301}, 164 (2013)
  [arXiv:1204.6333 [hep-ph]];
  D.~Liu, I.~Low and C.~E.~M.~Wagner,
  Phys.\ Rev.\ D {\bf 96}, no. 3, 035013 (2017)
  [arXiv:1703.07791 [hep-ph]];
  A.~Banerjee, G.~Bhattacharyya, N.~Kumar and T.~S.~Ray,
  JHEP {\bf 1803}, 062 (2018)
  [arXiv:1712.07494 [hep-ph]],
and references therein.

\end{thebibliography}
\end{document}